\documentclass[draft]{agujournal2019}
\usepackage{url} 
\usepackage[inline]{trackchanges} 
\usepackage{soul}
\usepackage{booktabs}
\usepackage{amsmath} 
\usepackage{amsmath,amssymb}

\draftfalse

\journalname{arXiv preprint}
\justifying

\begin{document}

%
%


\title{Toward Trustworthy Earthquake Catalogs in the Era of Automated Detection: A Probabilistic Framework for Robust Earthquake Location}

%
%




\authors{
        Ziye Yu\affil{1}, 
        Jinqing Sun\affil{1}, 
        Yuqi Cai\affil{1}, 
        Zemin Liu\affil{1}, 
        Pingping Wu\affil{1}, 
        Xin Liu\affil{2,3}, 
        Jiayan Tan\affil{1}}

\affiliation{1}{Institute of Geophysics, China Earthquake Administration,Beijing 100081, China}
\affiliation{2}{Laboratory of Seismology and Physics of Earth’s Interior, School of Earth and Space Sciences, University of Science and Technology of China, Hefei 230026, China}
\affiliation{3}{Institute of Advanced Technology, University of Science and Technology of China, Hefei 230088, China}





\correspondingauthor{Jinqing Sun}{yuziye@cea-igp.ac.cn}



\begin{keypoints}
\item We develop a fully probabilistic earthquake location framework with hierarchical robust modeling to quantify event credibility and construct statistically reliable catalogs.
\item A neural-network travel-time surrogate enables scalable Bayesian posterior sampling while preserving physically interpretable uncertainty estimates.
\item Uncertainty-based probabilistic screening reveals systematic underestimation of location errors in conventional workflows and improves catalog reliability without measurable loss of recall.
\end{keypoints}

%
%

%
%


\begin{abstract}
The rapid proliferation of deep-learning-based detection and phase association algorithms has dramatically increased the size and apparent completeness of automatically generated earthquake catalogs. However, automated processing inevitably introduces false detections, mis-associated arrivals, and poorly constrained events, rendering rigorous uncertainty quantification essential for ensuring catalog reliability.
We present a fully probabilistic earthquake location framework that jointly infers hypocenters, origin times, phase-dependent noise scales, and contamination levels within a unified Bayesian formulation. To robustly accommodate unreliable observations, we introduce a two-level hierarchical robustness strategy. At the first level, arrival-time residuals are modeled using a Student-$t$ scale-mixture representation, allowing moderate-to-large deviations to be adaptively down-weighted through latent scale variables. At the second level, we incorporate an explicit two-component contamination model in which each phase pick is probabilistically classified as either an inlier or an outlier via latent indicator variables, with phase-specific contamination rates inferred directly from the data. This dual-layer formulation disentangles heavy-tailed measurement noise from gross misdetections, thereby eliminating reliance on heuristic data rejection or manual thresholding.
Posterior sampling is coupled with a neural-network travel-time surrogate, enabling scalable and GPU-accelerated inference for large event sets while preserving physical consistency. Synthetic experiments demonstrate statistically calibrated posterior uncertainties, with empirical coverage closely matching true location errors. Application to the 2022 Luding $M_s$~6.8 aftershock sequence further shows that posterior uncertainty effectively identifies poorly constrained events; uncertainty-based screening reduces the catalog from 10{,}590 to 6{,}562 events without measurable degradation in recall.
By unifying hierarchical robust statistics with scalable Bayesian inference, the proposed framework provides a principled pathway toward constructing statistically trustworthy earthquake catalogs in the era of automated seismic monitoring.
\end{abstract}

\section*{Plain Language Summary}
Modern earthquake monitoring systems can detect many more earthquakes than in the past. While this improves catalog completeness, automated detection and phase picking also introduce timing errors, misidentified arrivals, and occasionally false events. As a result, some reported earthquake locations may appear precise but are actually poorly constrained. This makes it difficult to confidently interpret fault structures, study aftershock patterns, or assess seismic hazard.
In this study, we develop a new earthquake location method that explicitly accounts for unreliable data. Instead of treating all timing measurements as equally trustworthy, our approach allows for occasional large errors and even incorrect phase picks. The method automatically reduces the influence of questionable observations while still using the reliable information contained in the data. This leads to earthquake locations and uncertainty estimates that more accurately reflect data quality and network geometry.
To make this uncertainty-based analysis practical for large data sets, we use a deep-learning model to rapidly approximate three-dimensional seismic travel times. The neural network does not replace physical modeling; rather, it accelerates the repeated calculations required in a fully probabilistic framework. By combining robust statistical modeling with efficient 3-D travel-time prediction, our approach produces more conservative and physically meaningful earthquake locations, providing a clearer and more reliable picture of seismic activity.

%
%

%


%
%
%
%

\section{Introduction}

Earthquake catalogs constitute the foundational data set for seismological research, underpinning investigations of seismic hazard, fault geometry, earthquake source physics, and spatiotemporal seismicity patterns. The scientific value of a catalog depends not only on its completeness, but also critically on the reliability of hypocenter estimates and the realism of their associated uncertainty quantification. Over the past decade, advances in waveform-based phase picking and automated event detection have dramatically increased the number of earthquakes identified by local and regional seismic networks. In particular, deep-learning-based phase pickers and sequence-based association algorithms have enabled the detection of vast numbers of microearthquakes that were previously undetected, especially during dense aftershock sequences and swarm activity \cite{ross_generalized_2018,mousavi_earthquake_2020,zhang_loc-flow_2022, zhou_earthquake_2022}. In several well-instrumented regions, catalog sizes have expanded by more than an order of magnitude. This rapid growth provides unprecedented opportunities for high-resolution imaging of fault structures and for improved understanding of earthquake triggering and cascading processes.

However, the rapid expansion of automated earthquake catalogs introduces a fundamental challenge. Fully automated detection and association pipelines inevitably generate a non-negligible fraction of erroneous phase picks \cite{cai_deep_2025, cai_seismicxm_2026, mousavi_earthquake_2020, wang_deep_2019, zhu_ustc-pickers_2023}, misassociated arrivals, and poorly associated events \cite{ross_phaselink_2019, mcbrearty_pairwise_2019, yang_simultaneous_2021, zhang_rapid_2019, zhu_earthquake_2022, yu_fastlink_2022}. Even when such problematic observations represent only a small proportion of the data, their influence can disproportionately degrade earthquake location accuracy and bias downstream seismological inferences. Consequently, constructing trustworthy earthquake catalogs—catalogs that are not merely large, but scientifically reliable and accompanied by uncertainty estimates that faithfully reflect data quality—has become increasingly difficult. Robust and computationally scalable earthquake location methodologies are therefore required to keep pace with modern detection capabilities, while explicitly accommodating the imperfections inherent in automated seismic processing.

Traditional earthquake location techniques face fundamental limitations when applied to large, automatically generated datasets. Classical approaches based on iterative linearized least-squares inversion, such as Geiger-type methods \cite{geiger_probability_1912}, typically assume independent, Gaussian-distributed arrival-time errors and employ simplified velocity models. In automated detection environments, these assumptions are frequently violated. Arrival-time residuals may be strongly non-Gaussian, correlated across stations, or contaminated by misassociated phases, while unmodeled three-dimensional velocity heterogeneity can introduce systematic travel-time biases. As a consequence, formal uncertainty estimates derived under Gaussian assumptions may substantially underestimate true location errors.

Several approaches have been proposed to alleviate these problems. Relative relocation techniques, such as the double-difference algorithm \cite{waldhauser_double-difference_2000}, can significantly improve the relative geometry of earthquake clusters by suppressing common-mode errors. However, because these methods are formulated in terms of differential measurements, they do not directly correct absolute location bias and do not yield fully representative estimates of absolute location uncertainty.

In parallel, machine-learning-based approaches have been developed for earthquake location, including single-station methods \cite{ochoa_gutierrez_fast_2018, lockman_single-station_2005, zhu_machine_2024, munchmeyer_earthquake_2021} and end-to-end models that directly predict event locations from waveform data \cite{yang_simultaneous_2021, yano_graphpartitioning_2021, yoma_end--end_2022}. Although these methods can achieve high computational efficiency and, in some cases, competitive accuracy, most lack a probabilistic framework for event-level reliability assessment and statistically calibrated uncertainty quantification. Consequently, the credibility of individual events in large, automatically generated catalogs remains difficult to evaluate.

Over the past two decades, probabilistic earthquake location methods have established a rigorous foundation for uncertainty quantification. Bayesian formulations treat hypocentral parameters as random variables and infer their posterior distributions conditioned on observed arrival times and prior information. Early implementations, most notably the \texttt{NonLinLoc} framework \cite{nolet_probabilistic_2000, meyers_earthquake_2014}, demonstrated that full three-dimensional posterior probability density functions (PDFs) can be resolved even in complex velocity models. These approaches naturally accommodate nonlinear parameter trade-offs \cite{martinsson_robust_2013} and reveal extended or multimodal uncertainty structures that are not captured by point-estimate methods.

Subsequent work introduced more realistic statistical descriptions of observational errors within Bayesian frameworks. Heavy-tailed likelihood models, such as the Student’s $t$ distribution, reduce sensitivity to outliers and produce more stable uncertainty estimates when arrival-time data are noisy or partially inconsistent. However, heavy-tailed formulations implicitly attribute all discrepancies to increased noise variance. They do not explicitly distinguish between moderately noisy measurements and grossly erroneous arrivals caused by false detections or incorrect phase associations. In large, automatically generated catalogs, this distinction becomes critical. A small number of severely inconsistent picks can inflate posterior uncertainty or destabilize inference, even when most observations remain informative. These limitations highlight the need for probabilistic frameworks that explicitly model the credibility of individual observations rather than relying solely on generic robustness.

Despite their conceptual advantages, fully probabilistic location methods remain computationally demanding. Bayesian inference requires repeated forward-model evaluations to adequately explore the posterior distribution, which becomes prohibitively expensive when travel times are computed using numerical ray tracing or wave-equation solvers in three-dimensional velocity models. Recent studies have addressed this bottleneck through machine-learning-based surrogate models for seismic wave propagation. Neural networks trained to approximate travel-time calculations can provide orders-of-magnitude acceleration relative to conventional solvers while maintaining high accuracy within their training domains \cite{smith_eikonet_2021, waheed_pinneik_2021, grubas_neural_2023, taufik_neural_2023, agata_physics-informed_2025}. When integrated into Bayesian inference schemes, such surrogates enable scalable exploration of complex and potentially multimodal posterior structures \cite{agata_physics-informed_2025, smith_hyposvi_2021}.

In this study, we integrate these methodological advances to develop a fully probabilistic and computationally efficient earthquake location framework for constructing trustworthy catalogs from automated detection pipelines. The proposed approach combines a Bayesian location engine with a neural-network-based surrogate for three-dimensional travel-time prediction. By replacing repeated numerical forward calculations with a trained neural surrogate, the computational cost per posterior sample is substantially reduced, enabling large-scale Bayesian inference for dense earthquake catalogs. Posterior inference is performed using a Metropolis--Hastings within Gibbs sampling scheme, allowing flexible incorporation of prior information, hierarchical modeling, and non-Gaussian likelihood formulations within a unified probabilistic framework.

Beyond conventional heavy-tailed robustness, the framework adopts a two-level hierarchical error model. Arrival-time residuals are first represented using a Student-$t$ scale-mixture formulation to accommodate moderately large deviations. In addition, grossly inconsistent arrivals arising from false detections or incorrect phase associations are explicitly modeled through a probabilistic contamination component that infers the credibility of individual phase picks during inference. This separation between heavy-tailed noise and discrete contamination prevents a small number of problematic observations from inflating overall posterior uncertainty or destabilizing inference. The method yields full posterior ensembles of hypocentral parameters for each event, from which physically interpretable uncertainty regions and quantitative measures of event reliability are derived.

We validate the framework using both synthetic experiments and a large real-world dataset from the 2022 $M_w$~6.6 Luding earthquake sequence in southwestern China, which produced several thousand aftershocks within a short time window. Synthetic tests demonstrate that the neural surrogate and Bayesian sampler recover known source parameters and produce statistically calibrated uncertainty estimates. Application to the Luding sequence shows that the proposed method achieves location accuracy comparable to established deterministic and probabilistic locators while retaining a larger number of well-constrained events in the final catalog. Together, these results indicate that explicitly modeling observational credibility, combined with surrogate-accelerated Bayesian inference, provides a practical pathway toward earthquake catalogs that are both highly complete and statistically reliable in the era of automated seismic monitoring.

\section{Method}
\subsection{Overview of the uncertainty-aware earthquake location framework}

\begin{figure}[htbp!]
\centering
\includegraphics[width=1.0\linewidth]{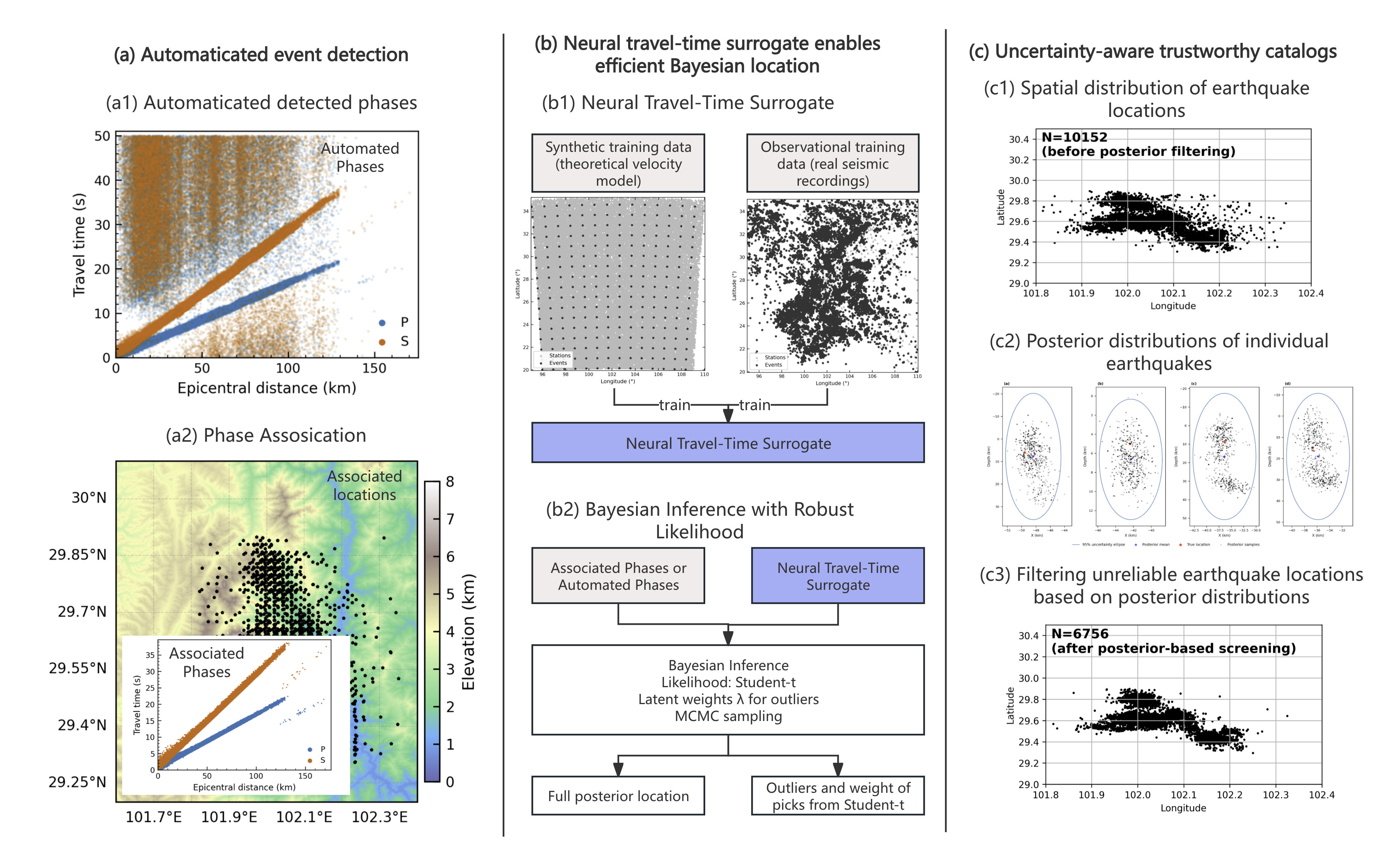}
\caption{\textbf{Overview of the uncertainty-aware framework for constructing trustworthy earthquake catalogs.}
(a) Automated phase picking and association.
(a1) Automatically detected P- and S-phase picks in epicentral distance–travel time space, illustrating scattered arrivals indicative of picking and association errors.
(a2) Spatial distribution of events after automated association, demonstrating the impact of erroneous picks on hypocenter estimates.
(b) Surrogate-accelerated Bayesian location.
(b1) A neural travel-time surrogate is trained using synthetic travel times generated from a three-dimensional velocity model and/or observational data under the same station geometry.
(b2) Bayesian inference integrates associated phase picks with the surrogate model. A hierarchical robust likelihood combines a Student-$t$ scale-mixture formulation with a probabilistic contamination component, and posterior distributions are sampled using Markov chain Monte Carlo (MCMC).
(c) Construction of uncertainty-aware catalogs.
(c1) Locations before posterior-based screening.
(c2) Representative posterior distributions illustrating event-specific uncertainty.
(c3) Locations retained after uncertainty-based screening, removing poorly constrained events and improving spatial coherence.}
\label{fig:workflow}
\end{figure}

Modern automated earthquake monitoring systems typically comprise three sequential stages: phase picking, phase association, and hypocenter estimation. Errors introduced during automated picking propagate through association into the location stage, where they may bias hypocenter estimates or generate spurious events. Figure~\ref{fig:workflow} summarizes the proposed uncertainty-aware framework, which is designed to robustly estimate earthquake locations, explicitly quantify posterior uncertainty, and construct statistically reliable catalogs.

The workflow begins with automated phase picking and association (Figure~\ref{fig:workflow}a). Deep neural network–based pickers identify candidate P- and S-phase arrivals, which are subsequently grouped into events through association algorithms. Although such pipelines substantially increase catalog completeness, residual picking errors and misassociations remain unavoidable. Even a small fraction of erroneous arrivals can disproportionately influence deterministic location procedures.

To enable scalable probabilistic inference for large catalogs, we introduce a neural travel-time surrogate $f_t(\cdot)$ to approximate seismic travel times in three-dimensional velocity structures (Figure~\ref{fig:workflow}b). The surrogate maps source–receiver coordinate pairs $(\mathbf{x}_s,\mathbf{x}_r)$ to P- and S-wave travel times,
\begin{linenomath*}
\begin{equation}
t_k = T_k(\mathbf{x}_s, \mathbf{x}_r), \quad k \in \{P,S\},
\label{eq:surrogate}
\end{equation}
\end{linenomath*}
where $T_k$ denotes the learned travel-time function. The surrogate may be trained using synthetic travel times generated from a three-dimensional velocity model or directly from observed data, without requiring explicit velocity inversion. Because posterior inference marginalizes over residual modeling uncertainty, strict physics-informed constraints are not imposed during training.

Earthquake locations are estimated within a Bayesian framework that integrates the neural surrogate with associated phase picks. The likelihood model adopts a two-level hierarchical error formulation. First, arrival-time residuals are represented using a Student-$t$ scale-mixture likelihood to accommodate moderately large deviations. Second, a probabilistic contamination component introduces latent indicator variables that explicitly model grossly inconsistent arrivals arising from false detections or incorrect phase associations. This separation between continuous heavy-tailed noise and discrete contamination prevents a small number of problematic observations from inflating posterior uncertainty or destabilizing inference.

Posterior sampling of hypocentral parameters is performed using a Metropolis--Hastings within Gibbs scheme, yielding full posterior ensembles rather than point estimates. From these ensembles, physically interpretable uncertainty regions and quantitative measures of event credibility are derived.

Finally, posterior uncertainty metrics are used to construct an uncertainty-aware earthquake catalog (Figure~\ref{fig:workflow}c). Events exhibiting diffuse or poorly constrained posterior distributions are identified and removed based on objective uncertainty criteria. This posterior-based screening step shifts catalog construction from heuristic residual filtering toward uncertainty-based probabilistic assessment, producing spatially coherent and scientifically reliable seismicity patterns suitable for downstream analyses.

\subsection{Probabilistic Earthquake Location with Robust Error Modeling}

We formulate earthquake location as a fully probabilistic inference problem and solve it using a Metropolis--Hastings-within-Gibbs sampling scheme. The framework jointly infers earthquake source locations, origin times, and phase-dependent uncertainty parameters for a set of events, while explicitly accounting for erroneous or inconsistent phase picks. All source and receiver coordinates are represented in a local projected Cartesian coordinate system, with distances expressed in kilometers.

For each event $e$ and phase pick $i$ assigned to that event, the observed arrival time is modeled as
\begin{linenomath*}
\begin{equation}
t^{\mathrm{obs}}_{k i}
=
t_{0,e} + T_k(\mathbf{x}^s_e, \mathbf{x}^r_i) + \epsilon_{k i},
\end{equation}
\end{linenomath*}
where $T_k(\cdot)$ denotes a neural-network-based travel-time surrogate for phase $k\in\{P,S\}$, $\mathbf{x}^s_e$ is the earthquake source location, $\mathbf{x}^r_i$ is the receiver location, and $t_{0,e}$ is the event origin time.

To achieve robustness against picking errors and unmodeled velocity heterogeneity, we adopt a two-component contamination model for the arrival-time residuals. Each phase pick is associated with a latent indicator $z_{k i}\in\{0,1\}$ that denotes whether the observation is treated as an inlier or an outlier. Conditional on $z_{k i}$, the residual $\epsilon_{k i}$ follows
\begin{linenomath*}
\begin{equation}
\begin{aligned}
\epsilon_{k i} \mid z_{k i}=1
&\sim \mathrm{Student\text{-}t}_{\nu_k}(0,\sigma^2_{k,e}), \\
\epsilon_{k i} \mid z_{k i}=0
&\sim \mathcal{N}(0,\sigma^2_{k,\mathrm{out}}),
\end{aligned}
\end{equation}
\end{linenomath*}
where $\sigma^2_{k,e}$ represents the event-specific inlier variance for phase $k$, and $\sigma^2_{k,\mathrm{out}}$ is a fixed, large variance representing grossly inconsistent or misassociated picks. The prior probability that a pick is an inlier is denoted by $\pi_k$, which is shared across all events for each phase.

The Student-$t$ component is implemented through its standard scale-mixture representation by introducing latent scale weights $\lambda_{k i}$, which adaptively down-weight large residuals. This formulation yields a heavy-tailed likelihood while preserving conditional conjugacy and computational efficiency.

We assign flat priors to the origin times $t_{0,e}$ and inverse-gamma priors to the inlier variances $\sigma^2_{k,e}$. Source locations are regularized using weakly informative priors, optionally specified as a mixture between a Gaussian mixture model and a broad isotropic Gaussian distribution. The inlier probabilities $\pi_k$ follow Beta priors, enabling data-driven estimation of the fraction of reliable phase picks for each seismic phase.

Let $\Theta=\{\mathbf{x}^s_e,t_{0,e},\sigma^2_{P,e},\sigma^2_{S,e}\}_{e=1}^{N_C}$,
$\Lambda=\{\lambda_{k i}\}$, $Z=\{z_{k i}\}$, and $\Pi=\{\pi_P,\pi_S\}$. The joint posterior density is given by
\begin{linenomath*}
\begin{equation}
\begin{aligned}
p(\Theta,\Lambda,Z,\Pi \mid \mathcal{D})
\;\propto\;
&\prod_{i,k}
p\!\left(
t^{\mathrm{obs}}_{k i}
\,\middle|\,
\mathbf{x}^s_{e(i)},\,t_{0,e(i)},\,\sigma^2_{k,e(i)},\,
\lambda_{k i},\,z_{k i}
\right)
\\[2pt]
&\times
\prod_{e=1}^{N_C}
p(\mathbf{x}^s_e)\,
p(t_{0,e})\,
p(\sigma^2_{P,e})\,
p(\sigma^2_{S,e})
\\[2pt]
&\times
\prod_{i,k}
p(\lambda_{k i})\,
p(z_{k i}\mid \pi_k)
\\[2pt]
&\times
\prod_{k\in\{P,S\}}
p(\pi_k),
\end{aligned}
\end{equation}
\end{linenomath*}
where $\mathcal{D}$ denotes the set of observed phase picks, missing phases are omitted via masking, and $N_C$ is the number of earthquakes.

Posterior sampling is performed using an MH-within-Gibbs scheme that alternates between:  
(i) updating latent scale weights $\lambda_{k i}$ and inlier indicators $z_{k i}$,  
(ii) updating the inlier probabilities $\pi_k$,  
(iii) updating source locations $\mathbf{x}^s_e$ using random-walk Metropolis-Hastings proposals, and  
(iv) updating origin times and inlier variances using closed-form conditional distributions.  
All updates are vectorized across events, enabling efficient parallel inference for large earthquake catalogs.

Proposal step sizes for source locations are adapted during an initial tuning phase to target stable acceptance rates and are then fixed. Only observations classified as inliers contribute to the updates of origin times and variance parameters, ensuring that grossly inconsistent picks do not bias location estimates. After discarding burn-in samples and applying thinning, the retained samples provide an approximation to the full posterior distribution of hypocenter parameters and associated uncertainties.

Overall, this probabilistic framework yields uncertainty-aware earthquake locations and provides an explicit, data-driven mechanism for identifying and down-weighting unreliable observations, enabling the construction of trustworthy earthquake catalogs from large, automatically detected datasets.

\subsection{Training Data and Velocity Model}

The neural travel-time surrogate was trained using synthetic first-arrival P- and S-wave travel times computed from a three-dimensional crustal velocity structure. We adopted the Chuan--Dian (Sichuan--Yunnan) crustal velocity model (version~2.0) developed by \citeA{liu_high-resolution_2023}, which provides gridded $V_P$ and $V_S$ fields derived from regional seismic tomography. This model serves solely as a physically realistic reference medium for generating synthetic travel times.

To ensure uniform numerical resolution and compatibility with the fast marching method (FMM) \cite{sethian_fast_1996}, the original velocity model was interpolated onto a regular Cartesian grid with 5~km spacing in both horizontal and vertical directions. The horizontal interpolation domain extends one degree beyond the original model bounds in both longitude and latitude, resulting in projected coordinate ranges of approximately $x\in[-685,685]$~km and $y\in[-815,850]$~km. The vertical extent spans depths from 0 to 70~km.

For computational efficiency in solving the eikonal equation, both earthquake sources and seismic stations were restricted to velocity grid nodes. At each horizontal grid location, 5–8 synthetic earthquakes were generated with depths randomly sampled between 0 and 50~km. Each event included between 4 and 25 randomly selected P- and S-phase arrivals, subject to a maximum epicentral distance of 300~km and the requirement that at least one station recorded both phases. This strategy ensures dense coverage of source–receiver geometries while maintaining computational tractability.

It is important to emphasize that the neural surrogate is not a physics-informed neural network (PINN) and does not learn the velocity structure itself. Instead, it approximates the mapping from source–receiver geometry to travel time. Consequently, once trained, the surrogate requires only travel-time observations and source–receiver coordinates as inputs. The Bayesian inference framework explicitly accounts for observational noise and contamination; therefore, the same surrogate architecture can be trained directly on real observed first-arrival times without requiring access to the underlying velocity model, provided that the training data adequately span the relevant geometric domain.

\section{Results}

\subsection{Travel-time prediction accuracy and residual characteristics}

We first assess the predictive accuracy and generalization capability of the neural travel-time surrogate using both synthetic test data and independent real earthquake observations.

Synthetic test events were generated independently from the training dataset and distributed on a regular three-dimensional grid with horizontal spacing of 100~km and source depths between 0 and 45~km. At each grid node, 5–8 events were simulated to provide multiple source realizations. The maximum epicentral distance was restricted to 300~km, consistent with the training configuration, ensuring that test geometries remained within the surrogate’s interpolation domain while avoiding overlap with the training samples.

To evaluate performance on real observations, we used earthquake phase catalogs provided by the China Earthquake Networks Center (CENC), comprising events that occurred in 2022 within the longitude range 97.0$^\circ$–108.0$^\circ$ and latitude range 21.0$^\circ$–34.0$^\circ$. The dataset includes 765{,}813 manually reviewed P- and S-phase picks, providing a large and high-quality benchmark for assessing surrogate accuracy under realistic observational conditions.

\begin{figure}[htbp!]
\centering
\includegraphics[width=0.8\linewidth]{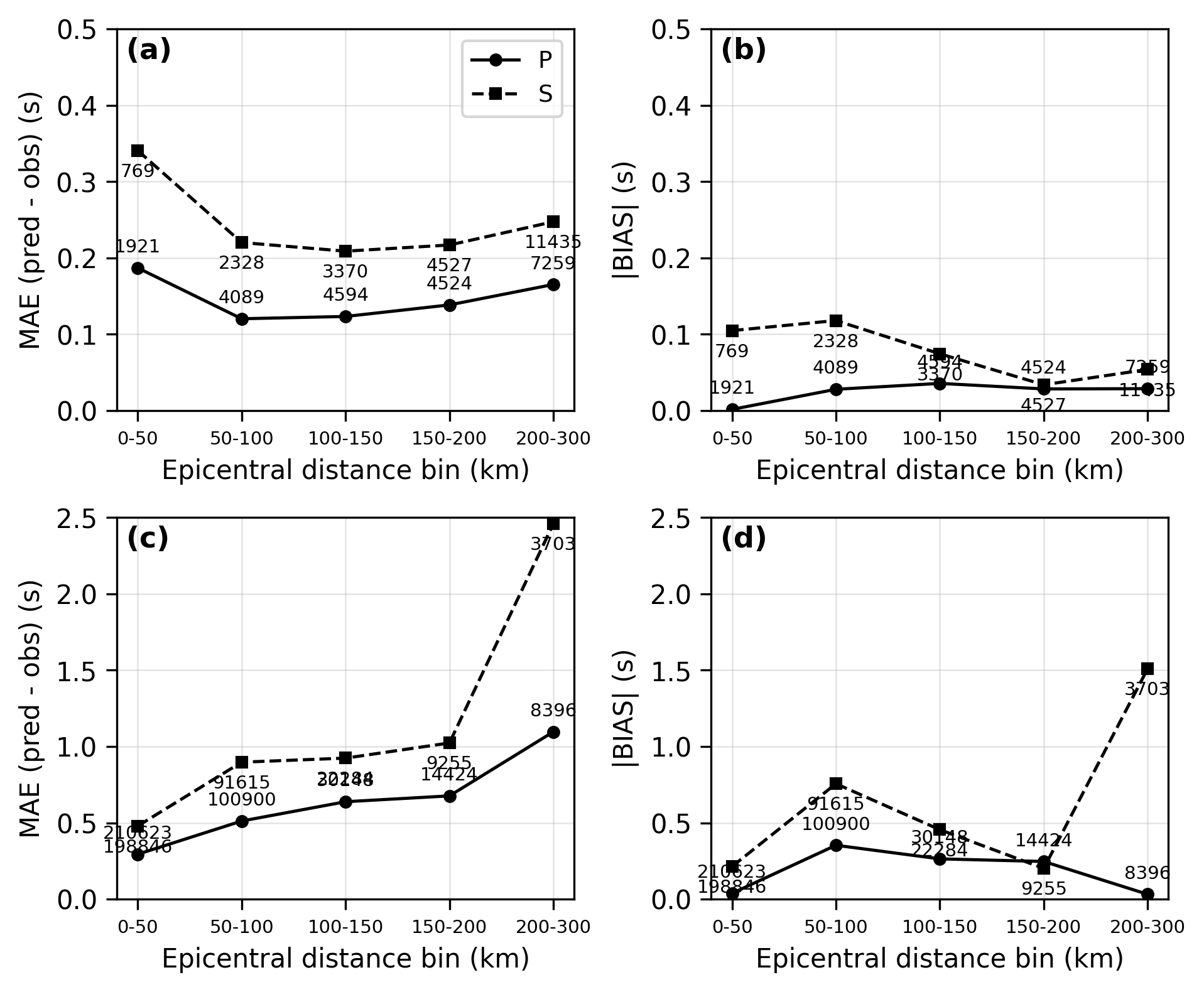}
\caption{
Performance of neural--network travel--time predictions for P and S phases.
(a) Mean absolute error (MAE) of predicted travel times as a function of epicentral distance bin.
(b) Absolute bias ($|\mathrm{BIAS}|$), defined as the absolute mean of travel--time residuals (predicted minus observed).
(c--d) Distributions of travel--time residuals for P and S phases, respectively, where positive values indicate delayed predictions.
Solid and dashed curves in panels (a) and (b) denote P and S phases, respectively.
Numbers adjacent to symbols indicate the number of picks in each distance bin.
Residual histograms are shown using identical binning, with the count axis displayed in scientific notation.
}
\label{fig:tt_accuracy}
\end{figure}

Travel-time prediction errors were computed as residuals between surrogate-predicted travel times and reference values (FMM solutions for synthetic tests and catalog arrival times for real data). Summary statistics, including mean residuals and mean absolute error (MAE), are shown in Figure~\ref{fig:tt_accuracy}. These results quantify both the intrinsic approximation error of the surrogate and its performance under real observational variability.

Figure~\ref{fig:tt_accuracy} summarizes the mean absolute error (MAE), absolute bias, and residual distributions for P and S phases.

For the synthetic test dataset (Figure~\ref{fig:tt_accuracy}a–b), the surrogate exhibits consistently low prediction errors across epicentral distance bins. The P-wave MAE remains below approximately 0.20~s, while the S-wave MAE ranges between 0.20 and 0.35~s. Absolute bias for both phases is small ($<0.12$~s) and substantially lower than the corresponding MAE, indicating that residuals are dominated by random approximation error rather than systematic offsets. These results confirm that the neural surrogate accurately reproduces FMM-computed travel times within the training domain.

In contrast, prediction errors increase when evaluated using independent real observations (Figure~\ref{fig:tt_accuracy}c–d), particularly at larger epicentral distances. For P waves, the MAE increases gradually from approximately 0.3~s at short distances to about 1.1~s beyond 200~km. S-wave errors exhibit a more pronounced increase, with MAE exceeding 2~s at large distances. Absolute bias also increases with distance and is notably larger for S phases, reaching values greater than 1~s at far offsets.

The discrepancy between synthetic and real-data performance primarily reflects modeling mismatch and observational variability. Real arrival times incorporate unmodeled lateral heterogeneity, imperfect phase identification, variable picking quality, and potential systematic differences between the reference velocity model used for surrogate training and the true Earth structure. Importantly, the residual patterns do not exhibit distance-dependent oscillatory bias or systematic instability, suggesting that the surrogate remains numerically stable even when applied to real observations.

Overall, these results indicate that the intrinsic approximation error of the neural surrogate is small under controlled conditions, and that performance degradation on real data is largely attributable to physical and observational factors rather than surrogate instability. This distinction motivates the subsequent adoption of a hierarchical probabilistic location framework that explicitly models observational uncertainty and contamination.

\subsection{Spatial distribution of earthquake locations and effects of uncertainty screening}

\begin{figure}[!htbp]
\centering
\includegraphics[width=\textwidth]{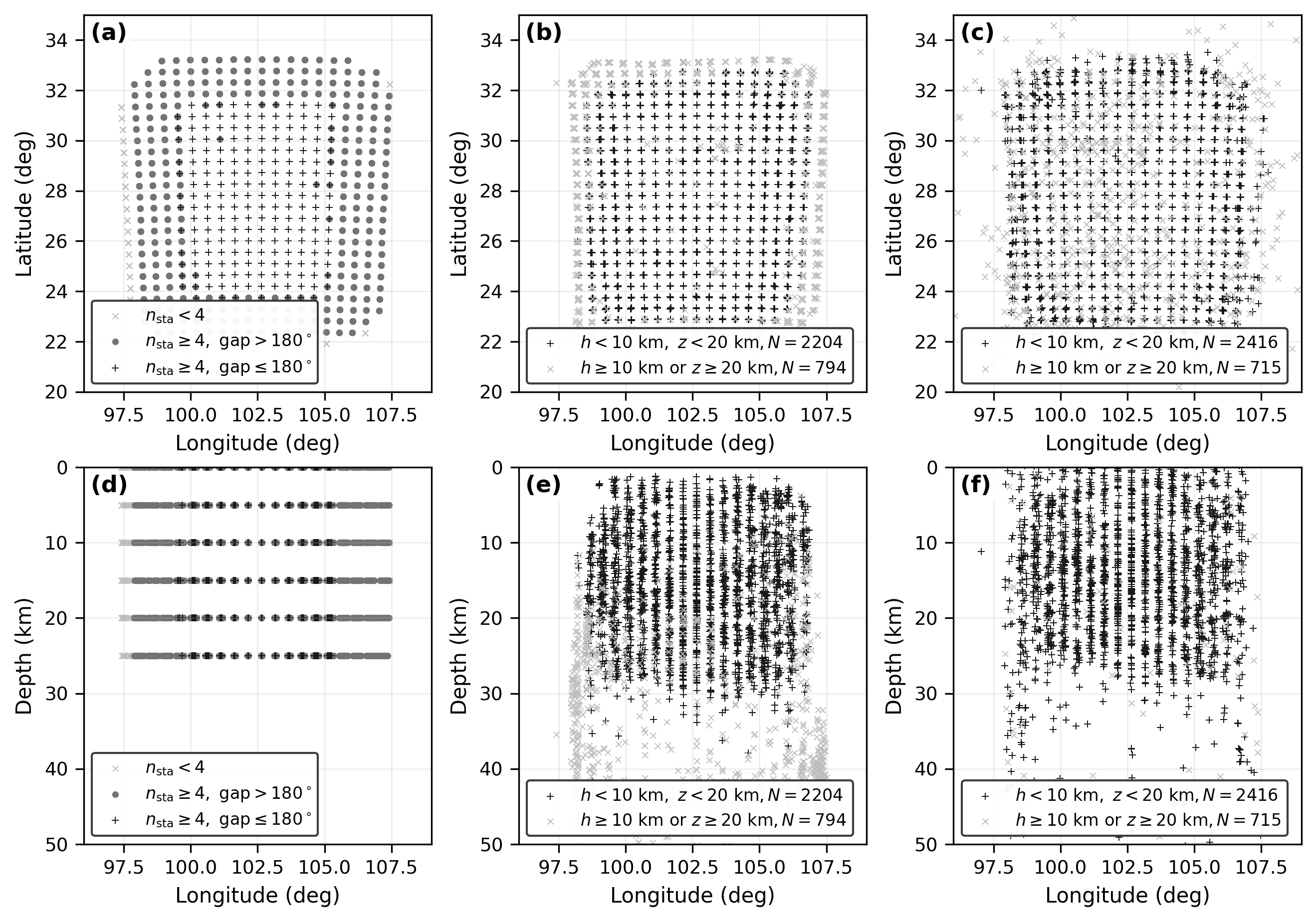}
\caption{
Spatial distribution of earthquake locations and associated quality metrics.
(a–c) Map view; (d–f) longitude–depth sections corresponding to the panels above.
(a,d) Reference catalog. Events are classified by the number of contributing stations ($n_{\mathrm{sta}}$) and azimuthal gap: gray circles indicate $n_{\mathrm{sta}} < 4$, light-gray symbols denote $n_{\mathrm{sta}} \ge 4$ with gap $>180^{\circ}$, and black crosses represent $n_{\mathrm{sta}} \ge 4$ with gap $\le 180^{\circ}$.
(b,e) Results from NLLoc. Events are separated according to horizontal uncertainty ($h$) and focal depth ($z$): black crosses indicate $h<10$~km and $z<20$~km ($N=2204$), whereas gray crosses denote $h\ge10$~km or $z\ge20$~km ($N=794$).
(c,f) Results from the proposed probabilistic method, using the same classification criteria: black crosses correspond to $h<10$~km and $z<20$~km ($N=2416$), and gray crosses indicate $h\ge10$~km or $z\ge20$~km ($N=715$).
}
\label{fig:location_compare}
\end{figure}

To evaluate location performance and uncertainty calibration, we generated an independent synthetic dataset (Figure~\ref{fig:location_compare}a,d). Hypocenters were placed on a regular 50~km horizontal grid with depths randomly sampled between 0 and 25~km, with 5–8 events simulated at each grid node. To introduce realistic variability in observational geometry, seismic stations were distributed up to 400~km from the network boundaries. For some events, fewer than four stations were available or azimuthal gaps exceeded $180^{\circ}$, intentionally creating unfavorable geometric conditions.

Figure~\ref{fig:location_compare} compares the reference catalog with results from NLLoc and the proposed probabilistic method. Events are classified using identical thresholds ($h<10$~km and $z<20$~km). In both methods, well-constrained events exhibit compact and spatially coherent distributions that broadly follow the reference pattern, whereas high-uncertainty solutions cluster near network edges where station coverage is sparse.

Uncertainty-based screening therefore primarily suppresses poorly constrained solutions associated with unfavorable geometry rather than correcting systematic location bias. Catalogs constructed using strict uncertainty thresholds appear spatially cleaner, but this compactness partly reflects the exclusion of high-uncertainty events.

The NLLoc results (Figure~\ref{fig:location_compare}b,e) display a particularly compact spatial pattern. This arises in part from its constrained search domain and predefined spatial bounds. Moreover, NLLoc travel times were computed on the same 5~km grid used to generate the synthetic dataset, ensuring forward-model consistency and near-optimal interpolation accuracy in this experiment. The comparison therefore does not disadvantage NLLoc in terms of travel-time representation fidelity; differences primarily reflect distinct statistical modeling strategies.

In contrast, the proposed probabilistic method yields a broader spatial distribution, especially near network boundaries, accompanied by a larger number of high-uncertainty events. Rather than imposing hard spatial constraints, the framework explicitly estimates posterior uncertainty by incorporating observational noise and station geometry within a hierarchical probabilistic model. As a result, weakly constrained events are retained but appropriately characterized by larger uncertainty.

Although the proposed method produces a visually less compact distribution than NLLoc, it provides a more transparent and internally consistent quantification of spatial uncertainty, which is essential for applications requiring rigorous uncertainty propagation.

Longitude–depth sections (Figure~\ref{fig:location_compare}e,f) further indicate that depth resolution is substantially weaker than horizontal resolution for both methods. Even under grid-consistent forward modeling, recovered focal depths exhibit noticeable dispersion relative to the reference distribution. This reflects the intrinsic trade-off between origin time and focal depth and the limited vertical resolving power of first-arrival travel times. The probabilistic framework makes this limitation explicit through broader posterior depth distributions rather than implicitly constraining depth variability.

\subsection{Geometric controls on location uncertainty}

\begin{figure}[!htbp]
\centering
\includegraphics[width=\textwidth]{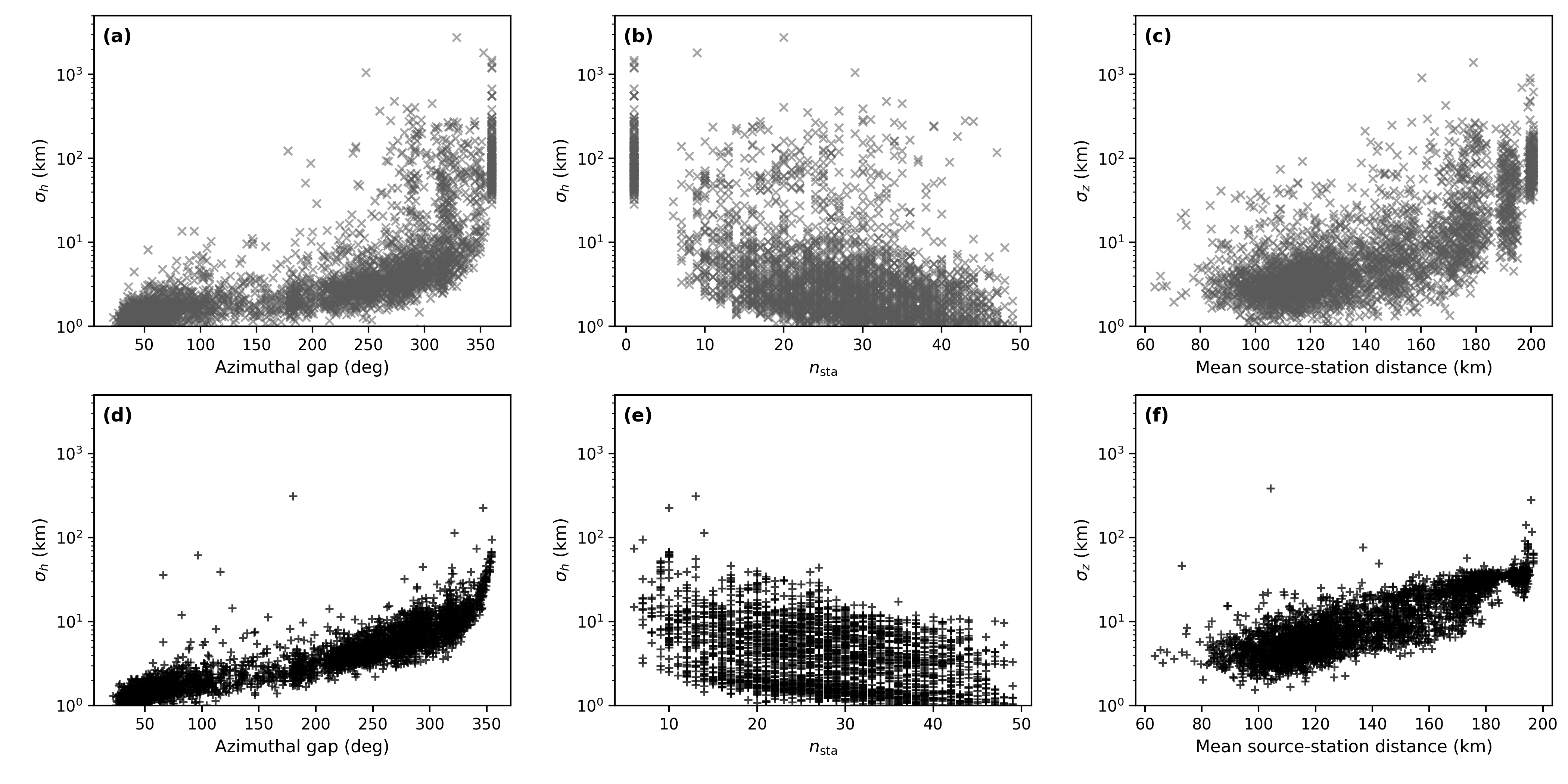}
\caption{Dependence of location uncertainty on geometric factors. Top row shows results from the proposed method, and bottom row shows results from NLLoc. (a, d) Horizontal uncertainty $\sigma_h$ versus azimuthal gap. (b, e) Horizontal uncertainty $\sigma_h$ versus number of contributing stations $n_{\mathrm{sta}}$. (c, f) Vertical uncertainty $\sigma_z$ versus mean source--station distance. All uncertainties are plotted on a logarithmic scale.}
\label{fig:geom}
\end{figure}

Figure~\ref{fig:geom} illustrates how estimated location uncertainties depend on key geometric factors, including azimuthal gap, number of contributing stations ($n_{\mathrm{sta}}$), and mean source–station distance. Results from the proposed probabilistic method are shown in the top row (Figures~\ref{fig:geom}a–\ref{fig:geom}c), and those from NLLoc in the bottom row (Figures~\ref{fig:geom}d–\ref{fig:geom}f). All uncertainties are plotted on a logarithmic scale to emphasize variability across orders of magnitude.

For both methods, horizontal uncertainty $\sigma_h$ increases with azimuthal gap and decreases with increasing station count, consistent with classical geometric resolution theory. Vertical uncertainty $\sigma_z$ increases with mean source–station distance in both approaches, reflecting the well-known degradation of depth resolution at larger epicentral distances.

Despite these common geometric trends, the magnitude and dispersion of uncertainties differ substantially. The proposed method produces a broad range of uncertainty values, with markedly larger estimates under poor azimuthal coverage or large epicentral distances. In contrast, NLLoc uncertainties remain relatively small and tightly clustered, even under unfavorable geometric conditions. This contrast indicates fundamentally different uncertainty propagation behaviors: the probabilistic framework amplifies uncertainty in response to degraded geometry, whereas NLLoc exhibits comparatively limited sensitivity to geometric deterioration.

\subsection{Posterior predictive checks based on travel-time residuals}
\begin{figure}[!htbp]
\centering
\includegraphics[width=\textwidth]{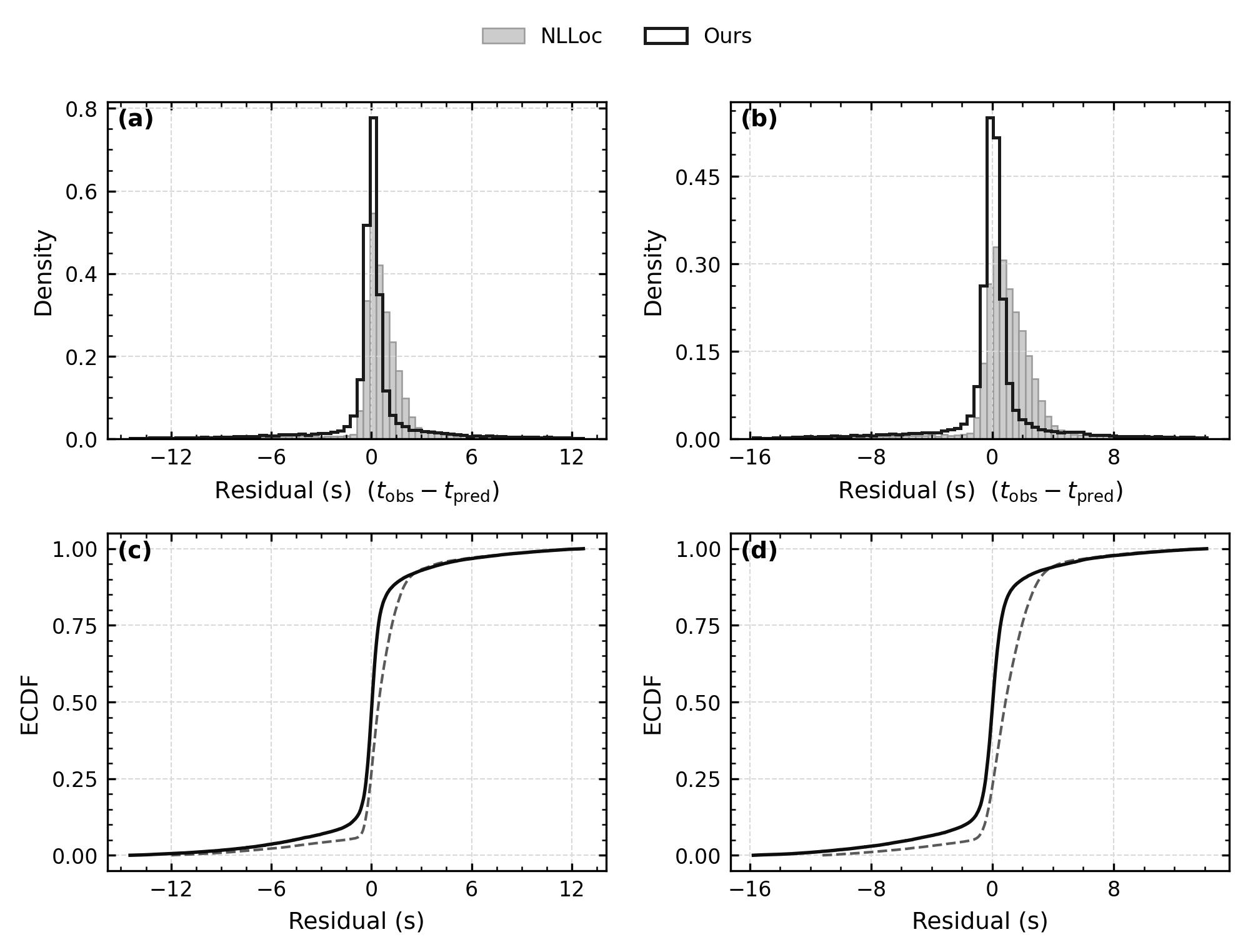}
\caption{Posterior predictive checks based on travel-time residuals. (a) Probability density functions of P-phase residuals. (b) Probability density functions of S-phase residuals. (c) Empirical cumulative distribution functions (ECDFs) of P-phase residuals. (d) ECDFs of S-phase residuals. Residuals are defined as observed minus predicted travel times using the PINN-based forward model.}
\label{fig:ppc}
\end{figure}

Figure~\ref{fig:ppc} presents posterior predictive checks based on P- and S-phase travel-time residuals, defined as the difference between observed arrivals and predictions from the neural travel-time surrogate. Probability density functions (Figures~\ref{fig:ppc}a–\ref{fig:ppc}b) and empirical cumulative distribution functions (ECDFs; Figures~\ref{fig:ppc}c–\ref{fig:ppc}d) are shown for both NLLoc and the proposed method.

For both phases, residuals from the proposed method are more strongly concentrated around zero, with a higher proportion of small-magnitude errors compared to NLLoc. This contrast is particularly evident in the ECDFs, where the probabilistic method attains higher cumulative probabilities at smaller residual thresholds. Both approaches exhibit heavy-tailed residual behavior, consistent with observational noise, modeling discrepancies, and structural complexity.

Uncertainty in the neural travel-time surrogate is not modeled as a separate stochastic component; instead, surrogate inaccuracies are implicitly absorbed into the heavy-tailed Student-$t$ likelihood. Explicit treatment of surrogate epistemic uncertainty remains an important direction for future development.

Overall, the posterior predictive checks demonstrate that the proposed method achieves a closer match between predicted and observed travel times, despite reporting larger nominal location uncertainties. The increased uncertainties therefore reflect a more conservative and internally coherent representation of predictive uncertainty rather than degraded data fit.

\subsection{Real Earthquake Experiments}

To evaluate the performance of the proposed earthquake location framework, we conducted a case study of the 2022 $M_w$~6.6 Luding earthquake sequence. Phase arrivals were automatically detected using the recurrent neural network (RNN) picker \cite{cai_deep_2025, yu_benchmark_2023} and subsequently associated into candidate events using the REAL algorithm \cite{zhang_rapid_2019}. The REAL association results reported by \citeA{liu2023auto}, which provide preliminary event locations and associated phase picks, were used as input to the proposed relocation framework.

\begin{figure}[!htbp]
\centering
\includegraphics[width=\linewidth]{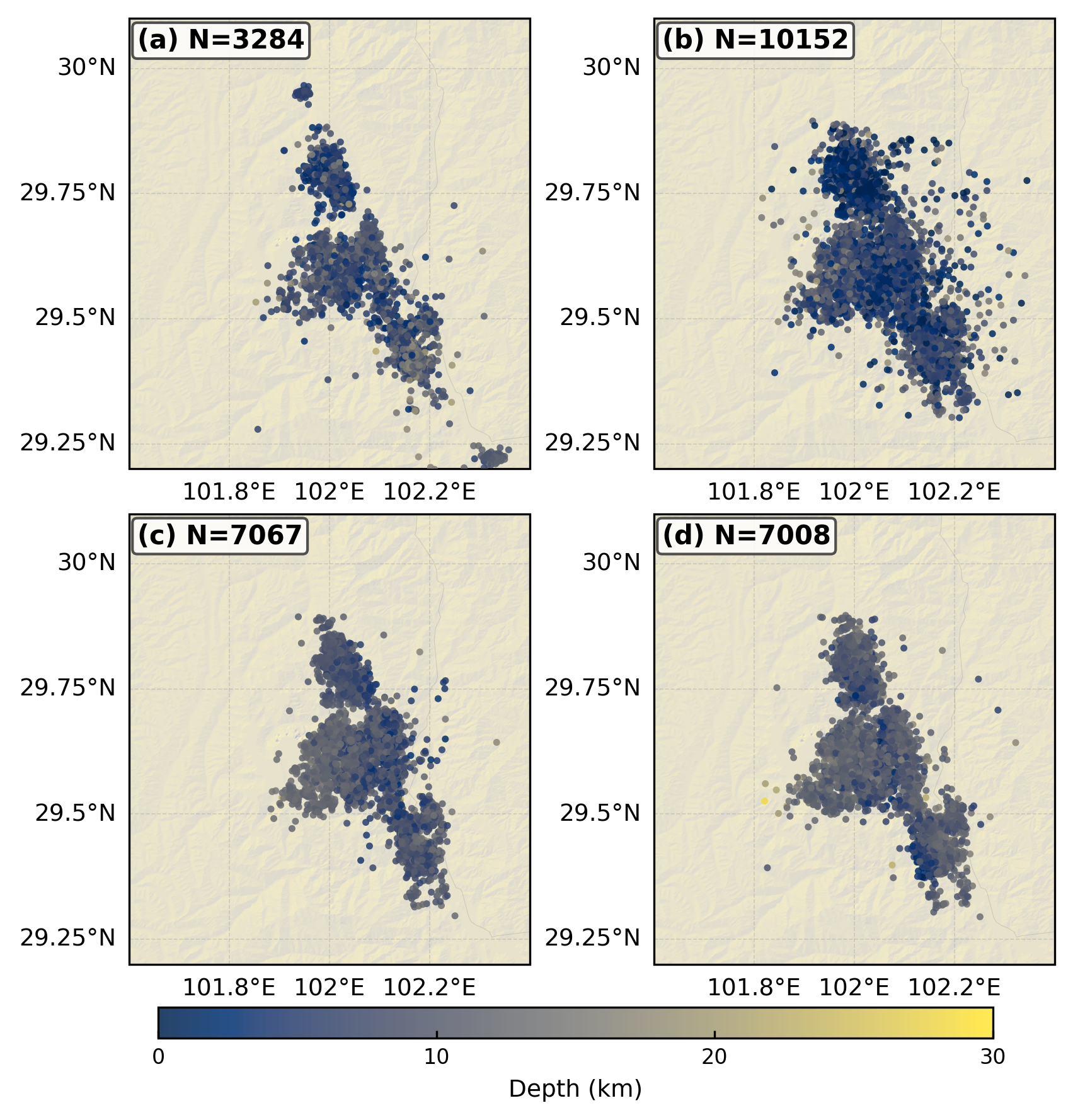}
\caption{
Spatial distributions of earthquake epicenters in the study area from four catalogs.
(a) The reference CENC catalog.
(b) The relocated catalog of Liu et~al.\ obtained using the Loc3D approach.
(c) Earthquake locations determined by NonLinLoc, after uncertainty-based screening with horizontal and vertical uncertainty thresholds of 3.2~km and 6.4~km, respectively.
(d) Results from the proposed method, screened using more permissive uncertainty thresholds of 10~km horizontally and 20~km in depth.
In all panels, earthquake epicenters are color-coded by focal depth, and the number of events retained after screening is indicated in the upper-left corner of each panel. A common color scale is used for all subplots to facilitate direct comparison.
}
\label{fig:luding_location}
\end{figure}

The resulting dataset spans 5–14 September 2022 and contains 3,284 events recorded by 53 stations within the geographic range 29.00$^{\circ}$–30.50$^{\circ}$N and 101.80$^{\circ}$–102.40$^{\circ}$E. The final relocation results are shown in Figure~\ref{fig:luding_location}.

To quantitatively assess catalog performance, we define a true positive (TP) event as one whose origin time differs from the reference catalog by less than 3~s and whose horizontal epicentral error is less than 20~km. Recall is defined as

\begin{linenomath*}
\[
\mathrm{Recall} = \frac{\mathrm{TP}}{\mathrm{TP} + \mathrm{FN}},
\]
\end{linenomath*}

where FN denotes false negatives relative to the reference catalog.

Using the REAL association results of \citeA{liu2023auto} as input, the overall recall of relocated events is 0.919 for all magnitudes and 0.873 for events with magnitude $\ge 3.0$.

\begin{figure}[!htbp]
\centering
\includegraphics[width=\linewidth]{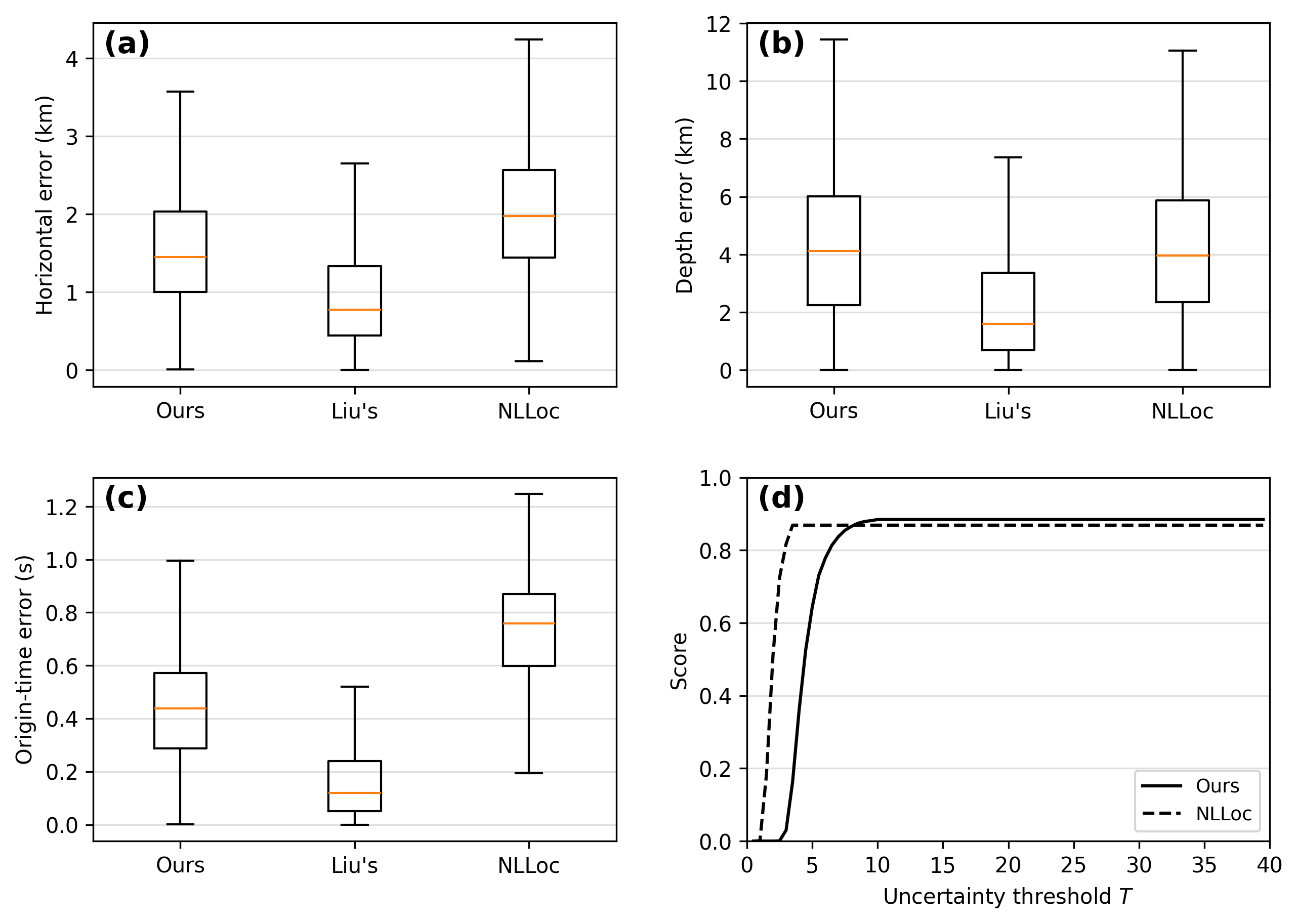}
\caption{Comparison of earthquake location accuracy and event recall for three location methods.
(a) Horizontal location error,
(b) depth error, and
(c) origin-time error for true positive events relative to the reference catalog. 
Boxes indicate the interquartile range, horizontal lines denote medians, and whiskers represent the full data range.
(d) Event recall as a function of the uncertainty threshold $T$, where events are retained only if the estimated horizontal uncertainty satisfies $e_h \le T$ and the vertical uncertainty satisfies $e_z \le 2T$.
The reference catalog is generated using Liu's method, which explains its smaller errors in panels~(a)--(c).
}
\label{fig:location_error}
\end{figure}

Figure~\ref{fig:location_error} compares the location accuracy of three earthquake location methods using true positive (TP) events identified relative to the reference catalog. Panels~(a)--(c) summarize the distributions of horizontal mislocation, depth mislocation, and origin-time error, respectively. For each event, errors are computed with respect to the reference catalog and statistics are shown using box-and-whisker plots.

As expected, the catalog-based solution implemented using Liu's method exhibits the smallest overall errors in all three metrics. This is primarily because the reference network catalog itself is generated using the same location framework, resulting in an internally consistent solution. Our method yields slightly larger but comparable errors, indicating that the proposed approach achieves a level of accuracy close to that of the operational catalog while relying on a fundamentally different inference strategy. In contrast, the NonLinLoc results show systematically larger median errors and broader distributions, particularly for depth and origin-time estimates.

Panel~(d) illustrates the event recall as a function of the uncertainty threshold $T$, where events are retained only if their estimated horizontal uncertainty satisfies $e_h \le T$ and the vertical uncertainty satisfies $e_z \le 2T$. For small thresholds, the recall of NLLoc increases rapidly, reflecting its tendency to underestimate formal uncertainties. Consequently, applying a stricter uncertainty threshold is sufficient to recover a large fraction of events. In comparison, the recall curve of our method increases more gradually with $T$, suggesting that the estimated uncertainties are better aligned with the actual location errors and provide a more reliable basis for quality control.

To further quantify the performance differences among the three location methods, we summarize the statistical results under fixed uncertainty thresholds in Table~\ref{tab:loc_benchmark}. For our method, events are retained using horizontal and vertical uncertainty thresholds of 15~km and 30~km, respectively, whereas a stricter set of thresholds (5~km and 10~km) is applied to the NonLinLoc (NLLoc) results to account for its tendency to underestimate formal uncertainties.

For all events, the three methods achieve comparable recall values (0.896--0.910), indicating that the majority of reference events can be successfully recovered under the adopted quality-control criteria. The catalog-based solution implemented using Liu's method exhibits the smallest mean horizontal, depth, and origin-time errors, which is expected because the reference catalog itself is generated using the same location framework. Our method yields slightly larger but comparable errors, while maintaining a recall level similar to that of the catalog solution. In contrast, NLLoc requires substantially tighter uncertainty thresholds to reach a comparable recall, yet still shows larger mean errors, particularly for horizontal location and origin time.

When considering only events with magnitude $M \ge 3$, the relative performance trends remain consistent. Although recall values decrease slightly for all methods due to the reduced sample size, our method and NLLoc achieve similar recall (0.857), while Liu's method shows a marginally lower value (0.841). In this magnitude range, Liu's method continues to provide the smallest horizontal errors, whereas our method yields intermediate errors and NLLoc exhibits larger depth and timing uncertainties. These results reinforce the conclusions drawn from Figure~\ref{fig:location_error}, demonstrating that the uncertainty estimates produced by our method are more consistent with actual location errors and therefore allow the use of physically meaningful quality-control thresholds.

\begin{table}[!htbp]
\centering
\caption{Performance comparison against the reference (manual) catalog.}
\label{tab:loc_benchmark}
\begin{tabular}{lccccc}
\hline
Method 
& Recall 
& \shortstack{Horizontal \\ error (km)} 
& \shortstack{Depth \\ error (km)} 
& \shortstack{Origin-time \\ error (s)} 
& Num. \\
\hline
\multicolumn{6}{l}{\textbf{All events}} \\
\hline
Ours\\{\footnotesize (screening: $h<10$~km, $z<20$~km)}
& 0.882 & 1.75 / 1.64 & 4.27 / 2.66 & 0.44 / 0.25 & 7008 \\

NonLinLoc\\{\footnotesize (screening: $h<3.2$~km, $z<6.4$~km)}
& 0.837 & 2.17 / 1.49 & 4.19 / 2.37 & 0.73 / 0.24 & 7067 \\

Liu et al.\ (2023)
& 0.893 & 1.17 / 1.62 & 2.37 / 2.31 & 0.19 / 0.26 & 10152 \\
\hline
\multicolumn{6}{l}{\textbf{Events with $M \ge 3$}} \\
\hline
Ours\\{\footnotesize (screening: $h<10$~km, $z<20$~km)}
& 0.857 & 1.48 / 0.78 & 2.92 / 2.22 & 0.38 / 0.36 & -- \\

NonLinLoc\\{\footnotesize (screening: $h<3.2$~km, $z<6.4$~km)}
& 0.571 & 1.63 / 0.51 & 3.00 / 1.76 & 0.70 / 0.24 & -- \\

Liu et al.\ (2023)
& 0.841 & 0.89 / 0.73 & 2.67 / 2.27 & 0.22 / 0.35 & -- \\
\hline
\end{tabular}

\vspace{0.5em}
\footnotesize{
\textit{Notes:} True-positive (TP) matches are defined by $|\Delta t| \le 3$~s and horizontal epicentral error $\le 20$~km.
Reported errors are absolute values and presented as mean / standard deviation.
Screening thresholds ($h/z$) denote horizontal and depth uncertainty limits applied for event retention.
}
\end{table}

\subsection{Inference Time Test}

\begin{table}[!htbp]
\centering
\caption{Comparison of computational performance between NonLinLoc and the proposed Bayesian location method.}
\label{tab:speed_compare}
\begin{tabular}{lcc}
\hline
Method 
& \shortstack{Total runtime \\ (s)} 
& \shortstack{Model size \\ (MB)} \\
\hline
NonLinLoc 
& 1153.2 
& 595.9 \\

Ours 
& 673.7 
& 4.8 \\
\hline
\end{tabular}

\vspace{0.5em}
\footnotesize{
\textit{Notes:} Experiments were conducted on an Apple M4 Max workstation. 
NonLinLoc was executed using 12 CPU cores, while the proposed method was accelerated using Apple Metal Performance Shaders (MPS). 
Reported runtime corresponds to full relocation of the same event dataset.
}
\end{table}

In practical seismic monitoring, computational efficiency is a critical consideration. In the proposed framework, travel-time evaluation is performed entirely by neural-network surrogates, allowing the full Bayesian inference pipeline to execute on GPU hardware without CPU-bound bottlenecks. The learned surrogate operates directly within a three-dimensional velocity model, thereby preserving 3-D travel-time heterogeneity while avoiding repeated numerical ray tracing.

This design leads to substantial computational gains. As summarized in Table~\ref{tab:speed_compare}, the end-to-end relocation of the test dataset requires 673.7~s using the proposed method, compared to 1153.2~s for NonLinLoc under 12-core CPU execution. In addition, the neural-network surrogate occupies only 4.8~MB, compared to 595.9~MB required for the precomputed travel-time tables used by NonLinLoc.

Overall, the proposed framework achieves a significant reduction in runtime and storage requirements while retaining full posterior uncertainty estimation. These improvements enhance the scalability of probabilistic earthquake location and make the method well suited for large-scale seismic monitoring and near–real-time applications.

\section{Discussion}

\subsection{Sensitivity to Initialization and Convergence Behavior}

\begin{figure}[htbp!]
    \centering
    \includegraphics[width=1.0\textwidth]{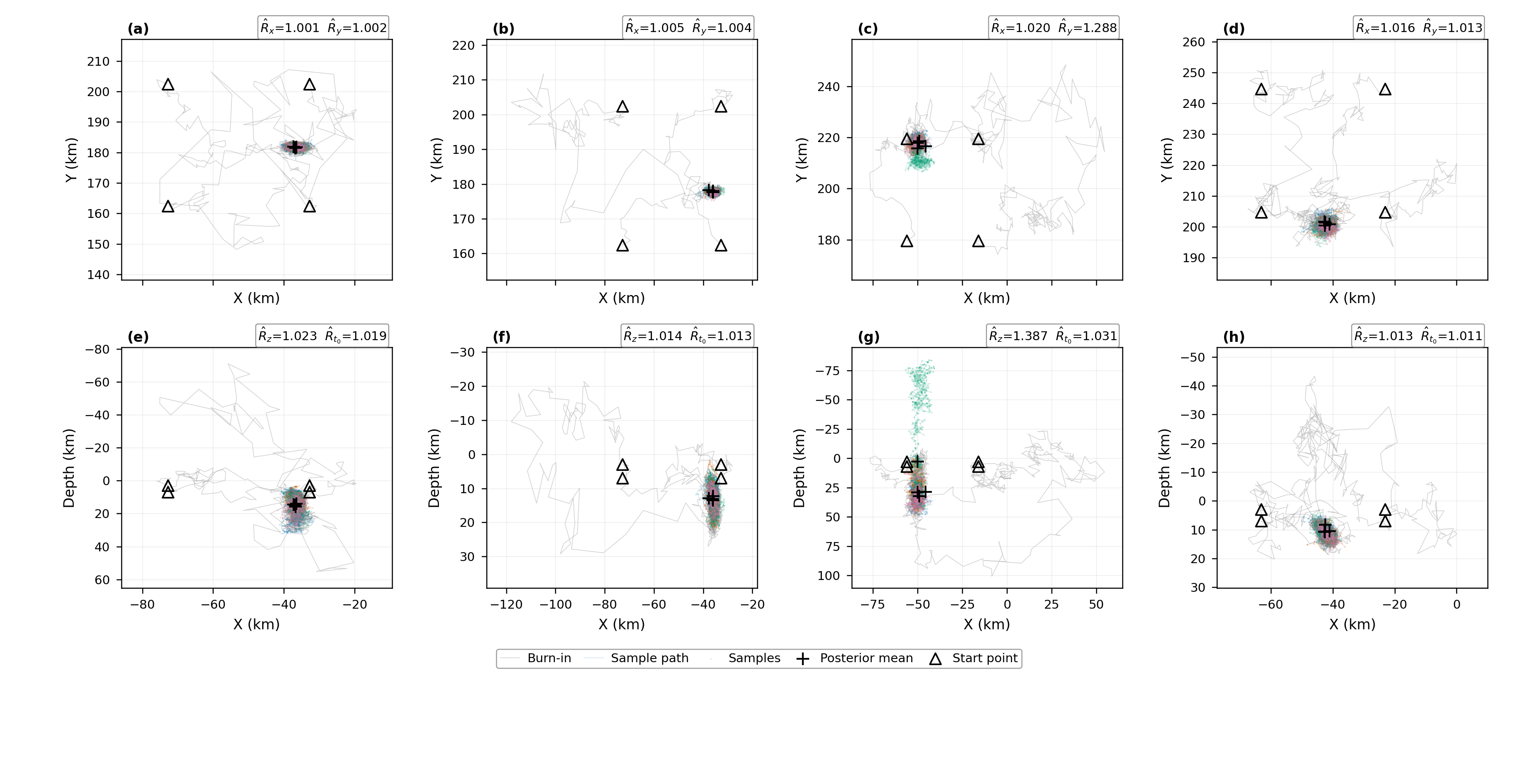}
    \caption{
    Sensitivity of MCMC-based earthquake location results to initialization for four randomly selected events.
    Panels (a)--(d) show horizontal projections of the sampling trajectories, and panels (e)--(h) show the corresponding depth--distance projections.
    For each event, four independent Markov chains are initialized from distinct starting locations with systematic offsets of $-5$, $0$, $+5$, and $+10$~km.
    Light gray curves indicate burn-in trajectories, colored points represent retained posterior samples, black crosses denote posterior means, and open triangles mark the initial locations.
    Despite large differences in the initial states, all chains converge to consistent high-probability regions.
    The Gelman--Rubin convergence diagnostics $\hat{R}$ for each parameter are reported in the panel headers, indicating satisfactory inter-chain convergence for both spatial coordinates and origin time.
    }
    \label{fig:mcmc_init_compare}
\end{figure}

The sensitivity of the proposed Bayesian earthquake location framework to initial model choices is evaluated by examining the convergence behavior of the Markov chain Monte Carlo (MCMC) sampler under deliberately perturbed initial hypocentral conditions.
Figure~\ref{fig:mcmc_init_compare} presents posterior sampling results for four randomly selected earthquake events, each initialized from four distinct starting locations with systematic offsets of $-5$, $0$, $+5$, and $+10$~km relative to a reference initial estimate.

Across all events, the burn-in trajectories (light gray curves) exhibit strong dependence on the imposed initial conditions, reflecting the expected behavior of random-walk MCMC samplers in high-dimensional and nonconvex posterior spaces typical of regional earthquake location problems.
In particular, the early-stage sampling paths may traverse large portions of the parameter space and display markedly different transient behaviors depending on the starting location.

Despite these pronounced differences during the burn-in phase, the retained posterior samples from all chains consistently converge toward compact and overlapping high-probability regions in both horizontal and vertical projections.
This behavior indicates that, for the selected events, the posterior distributions are well constrained by the arrival-time observations and are not dominated by the choice of initial hypocentral parameters.
The posterior means from different initializations (black crosses) cluster tightly within each panel, further confirming the stability of the inferred locations.

The convergence diagnostics support this visual assessment.
For all four events, the Gelman--Rubin statistics $\hat{R}$ for the horizontal coordinates remain close to unity (typically $\hat{R} < 1.02$), indicating satisfactory inter-chain mixing and convergence.
Similarly, the $\hat{R}$ values for depth and origin time are generally close to one, although slightly larger values are occasionally observed for depth, consistent with its weaker resolvability.
These diagnostics suggest that the chains have converged to a common stationary distribution despite the substantial differences in initialization.

It is important to emphasize that the convergence behavior illustrated here reflects the robustness of the sampling procedure and the identifiability of the posterior mode, rather than a direct quantification of location uncertainty.
While the horizontal posterior distributions are tightly clustered, the vertical distributions remain systematically broader, consistent with the well-known trade-offs between depth and origin time and the reduced sensitivity of arrival times to vertical perturbations.
Overall, these results demonstrate that the proposed Bayesian framework yields stable and reproducible hypocentral estimates and that its performance is not critically dependent on the choice of initial hypocenter for the examined events.

\subsection{Structural Sensitivity Analysis of the Location Framework}
\begin{table}[!htbp]
\centering
\caption{
Structural sensitivity analysis of relocation performance under alternative modeling configurations.
}
\label{tab:eikonal_benchmark}
\begin{tabular}{lccccc}
\toprule
Method & Recall & \shortstack{Horizontal \\ error (km)} & \shortstack{Depth \\ error (km)} & \shortstack{Time \\ error (s)} & Num. \\
\midrule
\multicolumn{6}{l}{\textbf{All events}} \\
\midrule
Ours \\ {\footnotesize (screening: 10 / 20 km)} 
& 0.882 & 1.75 / 1.64 & 4.27 / 2.66 & 0.44 / 0.25 & 7008 \\

Ours with Eikonal \\ {\footnotesize (screening: 13 / 26 km)} 
& 0.887 & 1.49 / 1.64 & 3.32 / 3.44 & 0.90 / 0.30 & 7232 \\

Ours without Latent $z$ \\ {\footnotesize (screening: 10 / 20 km)} 
& 0.878 & 1.74 / 1.61 & 4.30 / 2.52 & 0.44 / 0.25 & 6562 \\

Ours with real data \\ {\footnotesize (screening: 20 / 40 km)} 
& 0.846 & 3.77 / 1.30 & 6.21 / 8.71 & 0.38 / 0.26 & 6999 \\

Ours without elevation \\ {\footnotesize (screening: 10 / 20 km)} 
& 0.884 & 1.74 / 1.63 & 4.25 / 2.53 & 0.44 / 0.25 & 7035 \\
\midrule
\multicolumn{6}{l}{\textbf{Events with $M \ge 3$}} \\
\midrule
Ours \\ {\footnotesize (threshold: 10 / 20 km)} 
& 0.857 & 1.48 / 0.78 & 2.92 / 2.22 & 0.38 / 0.36 & -- \\

Ours with Eikonal \\ {\footnotesize (screening: 13 / 26 km)} 
& 0.857 & 1.01 / 0.65 & 3.25 / 2.39 & 0.70 / 0.41 & -- \\

Ours without Latent $z$ \\ {\footnotesize (screening: 10 / 20 km)} 
& 0.841 & 1.48 / 0.79 & 2.80 / 2.15 & 0.38 / 0.35 & -- \\

Ours with real data \\ {\footnotesize (screening: 20 / 40 km)} 
& 0.841 & 3.45 / 0.82 & 5.02 / 4.23 & 0.52 / 0.40 & -- \\

Ours without elevation \\ {\footnotesize (screening: 10 / 20 km)} 
& 0.857 & 1.59 / 1.19 & 3.02 / 2.25 & 0.39 / 0.36 & -- \\
\bottomrule
\end{tabular}

\vspace{0.5em}

\footnotesize{
True-positive (TP) matches are defined by $|\Delta t| \le 3$~s and horizontal epicentral error $\le 30$~km.
All reported errors are absolute values and presented as mean / standard deviation.
Screening thresholds refer to horizontal / depth posterior uncertainty limits applied for event retention.
}
\end{table}

We evaluated the structural sensitivity of the location framework under multiple alternative modeling configurations, including (1) Eikonal-based travel-time refinement, (2) removal of the latent contamination variable $z$, (3) omission of elevation corrections, and (4) substitution of the baseline travel-time surrogate with a model trained on regional catalog data (2009--2019). To ensure comparability across configurations, uncertainty-based screening thresholds were adjusted according to the posterior dispersion characteristics of each variant. Specifically, the baseline method employed horizontal and depth thresholds of 10/20~km, the Eikonal-refined model used 13/26~km, and the real-data-trained model used 20/40~km. Results are summarized in Table~\ref{tab:eikonal_benchmark}.

With adjusted thresholds, incorporation of Eikonal-based refinement yields a slight increase in recall (0.882 to 0.887 for the full event set) and moderate reductions in mean horizontal error (1.75~km to 1.49~km). However, these gains are accompanied by increased origin-time residuals and no systematic improvement in depth dispersion. For events with $M \ge 3$, recall remains unchanged (0.857), while horizontal accuracy improves but timing residuals increase. These results indicate that incremental refinements in deterministic forward modeling do not fundamentally alter posterior location behavior.

Removal of the latent contamination variable $z$ produces minimal changes in point-estimate errors but reduces recall, confirming that the contamination mechanism primarily stabilizes inference under observational inconsistency rather than directly enhancing geometric resolution.

In contrast, replacing the baseline travel-time model with one trained directly on real catalog data substantially degrades spatial accuracy, even under relaxed screening thresholds (20/40~km). The mean horizontal error increases to 3.77~km for the full event set, and depth dispersion also increases markedly. Although recall remains comparable for larger events, the overall deterioration in spatial consistency indicates that empirical travel-time regression alone does not guarantee physically coherent inverse solutions. This result underscores the importance of physically constrained forward modeling within the probabilistic framework.

Omission of elevation corrections produces negligible changes across all metrics, indicating limited sensitivity to this modeling component for the present dataset.

Collectively, these experiments demonstrate that posterior location performance within the Bayesian framework is governed primarily by station geometry, phase coverage, and intrinsic depth--origin-time trade-offs rather than by marginal refinements in deterministic travel-time prediction. The heavy-tailed Student-$t$ likelihood absorbs moderate forward-model discrepancies, while posterior sampling propagates timing uncertainty into spatial uncertainty, thereby limiting over-sensitivity to forward-model modifications.

Given the limited structural impact of forward-model refinement and the additional computational complexity introduced by such modifications, we retain the baseline travel-time formulation in the main workflow and treat alternative configurations as structural sensitivity tests.

\subsection{Precision versus geometric credibility in fault-related seismicity imaging}
\begin{figure}[htbp!]
\centering
\includegraphics[width=0.9\textwidth]{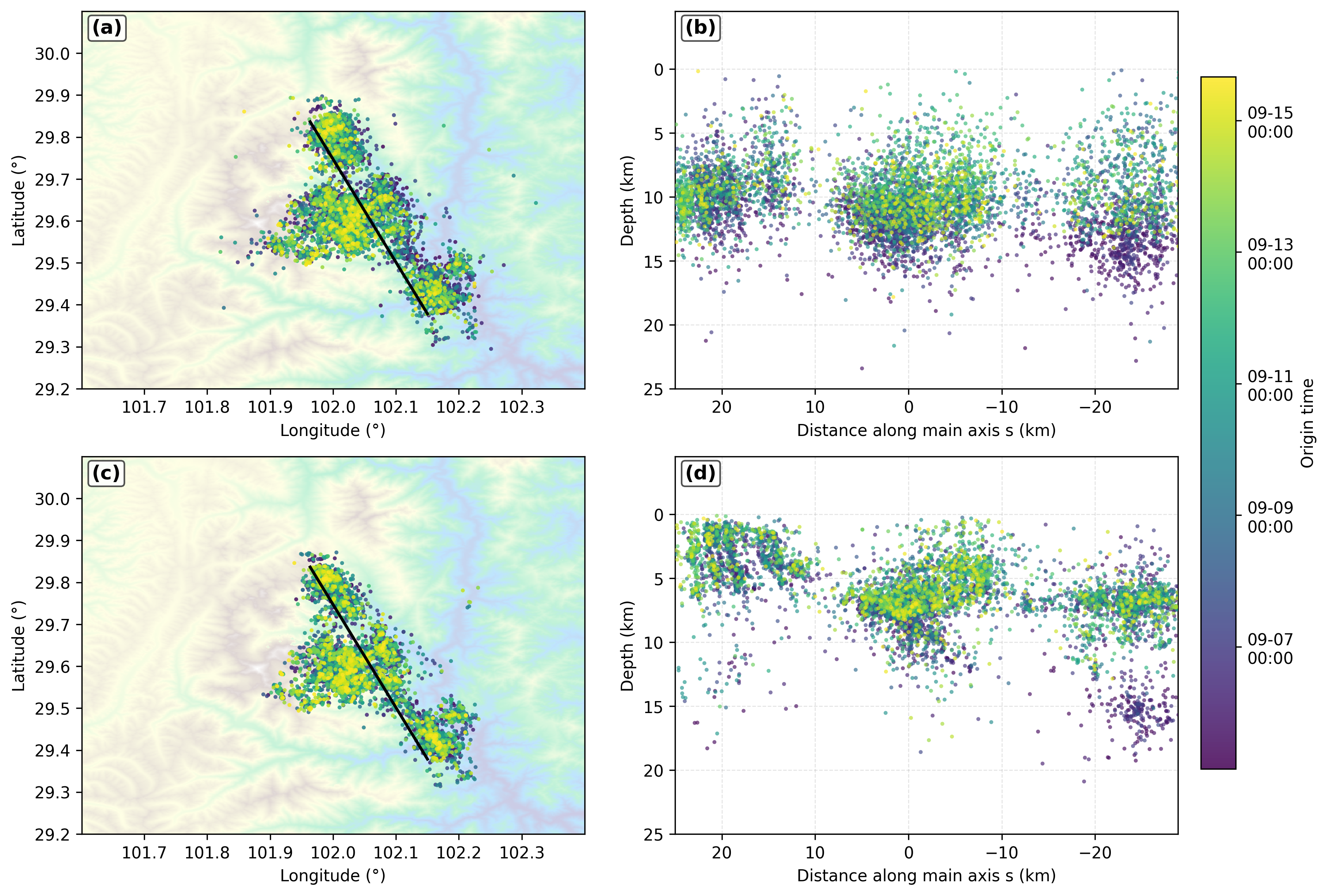}
\caption{Comparison of fault-related seismicity geometry obtained using our probabilistic location method (a,b) and HypoDD (c,d).
(a,c) Map views showing epicentral distributions colored by origin time; the black line indicates the principal axis of seismicity.
(b,d) Depth sections projected along the principal axis, with events within $\pm 10$~km of the axis displayed.
For each depth section, the median depth interquartile range (IQR) computed within 5-km along-strike bins is annotated.
Although HypoDD yields a smaller median depth IQR (2.21~km) than our method (2.85~km), the HypoDD results exhibit laterally continuous, near-horizontal depth bands indicative of depth compression by relative-location constraints.
In contrast, our probabilistic locations preserve finite fault-zone thickness and along-strike variability, consistent with heterogeneous resolution and physically plausible fault geometry.}
\label{fig:map_profile_2x2}

\end{figure}

A central motivation of this study is to examine how different earthquake location strategies influence the geometric characterization of fault-related seismicity, beyond simple improvements in formal location precision. Figure~\ref{fig:map_profile_2x2} compares our fully probabilistic absolute locations with results obtained using the relative-location algorithm HypoDD, highlighting differences in spatial compactness and depth variability.

In map view (Figures~\ref{fig:map_profile_2x2}a,c), both methods delineate a consistent overall strike of the dominant seismicity trend, indicating that the large-scale geometry is robust across approaches. However, differences emerge in the spatial concentration of events. Our probabilistic locations form a finite-width seismicity band aligned with the principal axis, whereas the HypoDD results exhibit tighter clustering around the main structure. This contrast reflects the distinct objectives of the two approaches: HypoDD is designed to minimize relative location residuals between nearby events, thereby enhancing relative precision, while our framework explicitly propagates observational uncertainty in absolute coordinates.

The contrast becomes more evident in the along-strike depth sections (Figures~\ref{fig:map_profile_2x2}b,d). The probabilistic solution displays moderate along-strike variability in both median depth and depth extent. In comparison, the HypoDD results appear more laterally coherent and vertically compact. We note that relative-location methods are generally more effective at resolving inter-event geometry than absolute depth when station coverage and takeoff-angle constraints are limited, which can lead to reduced apparent depth scatter.

To quantify these differences, we computed the depth interquartile range (IQR) within 5-km along-strike bins for events located within $\pm 10$~km of the principal axis. The median depth IQR is 2.85~km for our probabilistic locations and 2.21~km for HypoDD (Figures~\ref{fig:map_profile_2x2}b,d). The smaller IQR produced by HypoDD indicates higher relative compactness in depth. In contrast, the slightly larger IQR in our results reflects the propagation of observational uncertainty within the absolute-location framework.

These comparisons emphasize that spatial compactness alone does not fully characterize location quality when interpreting fault-zone geometry. Relative-location methods provide enhanced internal consistency, whereas probabilistic absolute approaches retain uncertainty-informed spatial variability. Together, the two perspectives offer complementary insights into the structure of the seismicity.

\subsection{Location with More Outlier Phases}

\begin{table}[!htbp]
\centering
\caption{Comparison of relocation performance against the reference (manual) catalog.}
\label{tab:raw_data}
\begin{tabular}{lccccc}
\hline
Method & Recall & \shortstack{Horizontal \\ error (km)} & \shortstack{Depth \\ error (km)} & \shortstack{Origin-time \\error (s)} & Num. \\
\hline
\multicolumn{6}{l}{\textbf{All events}} \\
\hline
Ours\\{\footnotesize (screening: $h<10$~km, $z<20$~km)}
& 0.822 & 1.72 / 1.60 & 4.36 / 2.62 & 0.45 / 0.24 & 5795 \\

Ours without Latent $z$\\{\footnotesize (screening: $h<10$~km, $z<20$~km)}
& 0.716 & 1.69 / 1.51 & 4.35 / 2.47 & 0.45 / 0.24 & 4611 \\

NonLinLoc\\{\footnotesize (screening: $h<3.2$~km, $z<6.4$~km)}
& 0.819 & 2.19 / 1.51 & 4.18 / 2.44 & 0.73 / 0.25 & 7405 \\
\hline
\multicolumn{6}{l}{\textbf{Events with $M \ge 3$}} \\
\hline
Ours\\{\footnotesize (screening: $h<10$~km, $z<20$~km)}
& 0.810 & 1.46 / 0.81 & 3.24 / 2.35 & 0.41 / 0.36 & -- \\

Ours without Latent $z$\\{\footnotesize (screening: $h<10$~km, $z<20$~km)}
& 0.778 & 1.47 / 0.81 & 3.24 / 2.18 & 0.40 / 0.37 & -- \\

NonLinLoc\\{\footnotesize (screening: $h<3.2$~km, $z<6.4$~km)}
& 0.571 & 1.59 / 0.53 & 2.99 / 1.68 & 0.70 / 0.24 & -- \\
\hline
\end{tabular}

\vspace{0.5em}
\footnotesize{
\textit{Notes:} True-positive (TP) matches are defined by $|\Delta t| \le 3$~s and horizontal epicentral error $\le 30$~km.
All reported errors are absolute values and presented as mean / standard deviation.
Screening thresholds ($h/z$) denote horizontal and depth uncertainty limits applied for event retention.
}
\end{table}

Because the proposed Bayesian framework explicitly models observational contamination through a latent outlier mechanism, it can, in principle, perform location without relying on conservative upstream association filtering. To evaluate this capability, we conducted an additional experiment in which phase picks rejected during the association stage were reintroduced into the dataset. In this setting, earthquake locations were inferred directly from the full set of associated and previously discarded picks, without additional association-based cleaning.

The resulting solutions were compared against the reference (manual) catalog using the same temporal and spatial matching criteria, and the performance is summarized in Table~\ref{tab:raw_data}. This experiment therefore tests robustness under elevated levels of observational contamination.

For the complete event set, the proposed method achieves a recall of 0.822, comparable to \texttt{NonLinLoc} (0.819) and substantially higher than the simplified model without the latent-$z$ mechanism (0.716). Horizontal, depth, and origin-time errors remain similar across methods, indicating that increased recall is not achieved at the expense of degraded precision. Notably, removal of the latent contamination variable $z$ leads to a marked reduction in recall while leaving location errors largely unchanged, demonstrating that the latent outlier model primarily improves event retention rather than point-estimate accuracy.

For events with $M \ge 3$, the contrast is more pronounced. The proposed framework maintains a recall of 0.810, whereas \texttt{NonLinLoc} drops to 0.571. Location errors remain comparable across methods, suggesting that the reduced recall of \texttt{NonLinLoc} reflects a higher sensitivity to contaminated or inconsistent phase picks rather than inferior geometric resolution.

These results highlight the role of the latent-$z$ contamination model. By explicitly inferring the credibility of individual phase observations, the Bayesian framework prevents a small number of erroneous picks from dominating event-level inference. Physically plausible events can therefore be retained even when some associated arrivals are inconsistent. In contrast, models lacking explicit contamination handling exhibit reduced robustness and a higher rate of effective event rejection under the same input conditions.

\subsection{Implications for trustworthy earthquake catalogs}

The differences illustrated in Figure~\ref{fig:map_profile_2x2} have direct implications for the construction of trustworthy earthquake catalogs intended for downstream analyses, such as fault mapping, seismicity clustering, and hazard assessment. The ability to quantify and propagate location uncertainty enables objective filtering or weighting of events based on posterior credibility rather than arbitrary distance thresholds. In this context, the preservation of realistic depth variability is essential, as artificially compressed catalogs may bias interpretations of seismogenic thickness, fault segmentation, and stress distribution.

More broadly, our results highlight the importance of distinguishing between statistical precision and physical credibility in earthquake location studies. Fully probabilistic approaches, when combined with efficient travel-time surrogates, make it feasible to retain this distinction at regional scales, offering a practical pathway toward earthquake catalogs that are not only precise, but also geologically meaningful.

\section{Conclusion}

We have presented a Bayesian earthquake‐location framework that couples a neural‐network surrogate for three‐dimensional travel‐time prediction with a Metropolis--Hastings‐within‐Gibbs sampling scheme to obtain full posterior distributions of hypocentral parameters. By replacing numerical ray tracing with a compact neural network trained on the CSNCD dataset, the forward modeling step becomes orders of magnitude faster while remaining consistent with the underlying 3-D velocity structure. The Bayesian formulation enables joint inference of location, origin time, and phase‐specific uncertainties, naturally yielding posterior probability distributions and credibility intervals that quantify the reliability of each estimate.

Application to the 2022 Luding earthquake sequence demonstrates that the proposed framework achieves location accuracy comparable to widely used deterministic and probabilistic approaches, including Loc3D and NonLinLoc, while improving computational efficiency by a factor of 10--50. The posterior samples delineate coherent seismicity patterns and reproduce the characteristic contrast between horizontal and vertical uncertainties. Incorporation of a Student-$t$ likelihood further enhances robustness to outliers, producing cleaner relocated catalogs and maintaining stable performance even under extreme-noise conditions where traditional algorithms degrade significantly.

These results underscore the potential of combining physics-informed machine learning with Bayesian inference to achieve real-time, uncertainty-aware seismic monitoring. The framework is computationally efficient, modular, and readily transferable to other regions or velocity models. Future developments may include joint inversion of velocity structure and hypocenters, tighter integration with automated phase-picking and association systems, and extensions to incorporate anisotropy or multi-arrival travel times. Overall, the method provides a practical pathway toward real-time probabilistic earthquake location in complex media and offers transparent uncertainty quantification essential for modern seismic hazard assessment.

\section*{Open Research}
The CSNCD dataset \cite{An2023_CSNCD} used for training and testing. 
The continuous waveform data used for the continuous-data evaluation are available from the International Earthquake Science Data Center \cite{IESDC}. 
The code (and the exact version used to produce the results) is archived \cite{yu_2026_location}.

\acknowledgments
The authors would like to thank Supported by the Deep Earth Probe and Mineral Resources Exploration - National Science and Technology Major Project of China (No. 2025ZD1007500) and the Special Fund of the Institute of Geophysics, China Earthquake Administration under Grant No. DQJB23R31. All scientific content was written and verified by the authors.

\section*{Conflict of Interest}
The authors declare no conflicts of interest relevant to this study.

%
%


%
%
%
%
%
\clearpage
\appendix

\section{The architecuture of travel time model}
The neural network $g_\theta$ maps the input vector $\mathbf{z}$ to a two-component output representing predicted P- and S-wave travel times:
\begin{equation}
    \hat{\mathbf{t}} = g_\theta(\mathbf{z}) = (\hat{t}_P, \hat{t}_S).
\end{equation}
The model architecture consists of seven hidden layers with 256 units each and hyperbolic tangent activations. A final linear layer followed by a sigmoid activation produces a two-dimensional output, which is scaled by a factor of 1000 to ensure that predicted travel times fall within physically plausible ranges:
\begin{equation}
    \hat{\mathbf{t}} = 
    1000 \, \sigma(\mathbf{W}_8 \mathbf{h}_7 + \mathbf{b}_8),
\end{equation}
where $\mathbf{h}_7$ denotes the activation of the final hidden layer and $\sigma(\cdot)$ is the logistic sigmoid function. This formulation provides a smooth and bounded regression output appropriate for regional travel-time prediction. A summary of the network architecture is given in Table~\ref{tab:traveltime_arch}.

\begin{table}[h]
\centering
\caption{Architecture of the neural network used for P- and S-wave travel-time regression.}
\label{tab:traveltime_arch}
\begin{tabular}{llll}
\hline
Layer & Type & Input $\rightarrow$ Output Size & Activation \\ \hline
Input & Concatenation & $(\mathbf{x},\mathbf{x}_s) \in \mathbb{R}^6$ & -- \\
Preprocessing & Scaling & $\mathbf{z} = [\mathbf{x},\mathbf{x}_s]/1000$ & -- \\
Hidden 1 & Fully connected & $6 \rightarrow 256$ & $\tanh$ \\
Hidden 2 & Fully connected & $256 \rightarrow 256$ & $\tanh$ \\
Hidden 3 & Fully connected & $256 \rightarrow 256$ & $\tanh$ \\
Hidden 4 & Fully connected & $256 \rightarrow 256$ & $\tanh$ \\
Hidden 5 & Fully connected & $256 \rightarrow 256$ & $\tanh$ \\
Hidden 6 & Fully connected & $256 \rightarrow 256$ & $\tanh$ \\
Hidden 7 & Fully connected & $256 \rightarrow 256$ & $\tanh$ \\
Output & Fully connected & $256 \rightarrow 2$ & SoftPlus $\times 10$ \\ \hline
\end{tabular}
\end{table}

\section{Details of the Bayesian Inference Framework}

\subsection{Bayesian Hypocenter Inference with Student-$t${t} Errors and Explicit Outlier Modeling}

We present the full Bayesian formulation underlying the probabilistic earthquake location framework described in the main text. Earthquake hypocenters are inferred using a Metropolis--Hastings-within-Gibbs (MH-within-Gibbs) sampler that jointly estimates source locations, origin times, phase-dependent uncertainty parameters, and latent variables that explicitly represent erroneous phase picks. The model combines a heavy-tailed Student-$t$ likelihood for inlier observations with a two-component contamination model that probabilistically separates reliable and grossly inconsistent arrivals.

All source and receiver coordinates are expressed in a local projected Cartesian coordinate system, with distances measured in kilometers and times in seconds.

We consider a catalog of $N_C$ earthquakes indexed by $e \in \{1,\dots,N_C\}$. Each phase pick $i \in \{1,\dots,N\}$ is associated with an event through an index map $e(i)$. For event $e$, the unknown parameters include the source location $\mathbf{x}^s_e \in \mathbb{R}^3$, the origin time $t_{0,e}$, and the phase-specific inlier variances $\sigma^2_{P,e}$ and $\sigma^2_{S,e}$. The receiver location corresponding to pick $i$ is denoted by $\mathbf{x}^r_i$. Seismic travel times are predicted by a neural-network-based forward model $T_k(\mathbf{x}^s,\mathbf{x}^r)$ for phase $k\in\{P,S\}$.

To explicitly account for erroneous observations, each phase pick is associated with a binary latent indicator $z_{k i}\in\{0,1\}$, where $z_{k i}=1$ denotes an inlier and $z_{k i}=0$ denotes an outlier. Grossly inconsistent picks are modeled using a fixed, large variance $\sigma^2_{k,\mathrm{out}}$. The prior distribution for the inlier indicator is
\begin{linenomath*}
\begin{equation}
z_{k i} \sim \mathrm{Bernoulli}(\pi_k),
\qquad
\pi_k \sim \mathrm{Beta}(a_\pi,b_\pi),
\end{equation}
\end{linenomath*}
where $\pi_k$ denotes the phase-dependent probability that a pick is reliable and is shared across all events.

The Student-$t$ inlier component is implemented using a standard scale-mixture representation. For inlier observations ($z_{k i}=1$), latent scale weights $\lambda_{k i}$ are introduced such that
\begin{linenomath*}
\begin{equation}
\begin{aligned}
\epsilon_{k i} \mid \lambda_{k i}, z_{k i}=1
&\sim \mathcal{N}\!\left(0,\frac{\sigma^2_{k,e(i)}}{\lambda_{k i}}\right),\\
\lambda_{k i}
&\sim \mathrm{Gamma}\!\left(\frac{\nu_k}{2},\frac{\nu_k}{2}\right),
\end{aligned}
\end{equation}
\end{linenomath*}
using the shape--rate parameterization. This representation yields conditionally Gaussian likelihoods and enables efficient Gibbs updates. Latent scale weights are defined only for inlier observations.

The remaining prior distributions are specified as
\begin{align}
p(t_{0,e}) &\propto 1,\\
\sigma^2_{k,e} &\sim \mathrm{Inv\text{-}Gamma}(\alpha_0,\beta_0),\\
p(\mathbf{x}^s_e)
&=
\pi_{\mathrm{mix}}\,p_{\mathrm{GMM}}(\mathbf{x}^s_e)
+
(1-\pi_{\mathrm{mix}})\,
\mathcal{N}\!\left(\mathbf{0},\sigma^2_{\mathrm{wide}}\mathbf{I}\right),
\end{align}
where the source-location prior can optionally incorporate a Gaussian mixture model (GMM) informed by prior geological or seismological information, with a broad isotropic Gaussian as a fallback. In this study, $\pi_{\mathrm{mix}}$ is set to zero.

\subsection{Conditional Posterior Updates}

For inlier observations, the conditional posterior distribution of the latent scale weights is
\begin{linenomath*}
\begin{equation}
\lambda_{k i}\mid r^{(k)}_i
\sim
\mathrm{Gamma}\!\left(
\frac{\nu_k+1}{2},
\frac{\nu_k+(r^{(k)}_i)^2/\sigma^2_{k,e(i)}}{2}
\right),
\end{equation}
\end{linenomath*}
where the residual is defined as
$r^{(k)}_i=t^{\mathrm{obs}}_{k i}-t_{0,e(i)}-T_k(\mathbf{x}^s_{e(i)},\mathbf{x}^r_i)$.

The conditional probability that a pick is an inlier is given by
\begin{linenomath*}
\begin{equation}
p(z_{k i}=1\mid\cdot)
=
\frac{\pi_k\,p_{\mathrm{in}}(r^{(k)}_i)}
{\pi_k\,p_{\mathrm{in}}(r^{(k)}_i)+(1-\pi_k)\,p_{\mathrm{out}}(r^{(k)}_i)},
\end{equation}
\end{linenomath*}
where $p_{\mathrm{in}}$ and $p_{\mathrm{out}}$ denote the inlier and outlier likelihood functions, respectively.

The phase-dependent inlier probabilities are updated according to
\begin{linenomath*}
\begin{equation}
\pi_k\mid Z
\sim
\mathrm{Beta}\!\left(
a_\pi+\sum_i\mathbb{I}[z_{k i}=1],
\;
b_\pi+\sum_i\mathbb{I}[z_{k i}=0]
\right).
\end{equation}
\end{linenomath*}

Conditioned on inlier observations only, the posterior distribution of the origin time for event $e$ is Gaussian with
\begin{linenomath*}
\begin{equation}
\begin{aligned}
\sigma^2_{t_0,e}
&=
\left(
\sum_{k}
\frac{1}{\sigma^2_{k,e}}
\sum_{i\in\mathcal{I}^{\mathrm{good}}_{k,e}}
\lambda_{k i}
\right)^{-1},\\
\mu_{t_0,e}
&=
\sigma^2_{t_0,e}
\sum_{k}
\frac{1}{\sigma^2_{k,e}}
\sum_{i\in\mathcal{I}^{\mathrm{good}}_{k,e}}
\lambda_{k i}
\bigl(t^{\mathrm{obs}}_{k i}-T_k(\mathbf{x}^s_e,\mathbf{x}^r_i)\bigr).
\end{aligned}
\end{equation}
\end{linenomath*}

The conditional posterior distribution of the inlier variances is
\begin{linenomath*}
\begin{equation}
\sigma^2_{k,e}
\sim
\mathrm{Inv\text{-}Gamma}\!\left(
\alpha_0+\frac{1}{2}\sum_{i\in\mathcal{I}^{\mathrm{good}}_{k,e}}\lambda_{k i},
\;
\beta_0+\frac{1}{2}\sum_{i\in\mathcal{I}^{\mathrm{good}}_{k,e}}
\lambda_{k i}(r^{(k)}_i)^2
\right).
\end{equation}
\end{linenomath*}

Each source location $\mathbf{x}^s_e$ is updated using a random-walk Metropolis--Hastings proposal,
\begin{equation}
\mathbf{x}^s_{e,\mathrm{new}}=\mathbf{x}^s_e+\boldsymbol{\eta}_e,
\qquad
\boldsymbol{\eta}_e\sim\mathcal{N}(\mathbf{0},s_e^2\mathbf{I}),
\end{equation}
with acceptance probability determined by the change in the joint log-posterior density. Event-specific proposal scales $s_e$ are adapted during an initial tuning phase to achieve stable acceptance rates and are subsequently held fixed.

\subsubsection{Sampling Procedure and Implementation Details}

Each MH-within-Gibbs iteration proceeds by sequentially updating
$\{\lambda_{k i}\}$, $\{z_{k i}\}$, $\{\pi_k\}$, $\{\mathbf{x}^s_e\}$,
$\{t_{0,e}\}$, and $\{\sigma^2_{k,e}\}$. Initial source locations are set to the centroid of associated stations with a small depth offset, and latent variables are initialized to favor inliers. Hyperparameters are fixed to weakly informative values. After discarding burn-in samples and applying thinning, the retained samples approximate the joint posterior distribution of earthquake hypocenters, origin times, uncertainty parameters, and inlier probabilities.

\section{Coverage and calibration of uncertainty estimates}
\begin{figure}[!htbp]
\centering
\includegraphics[width=\textwidth]{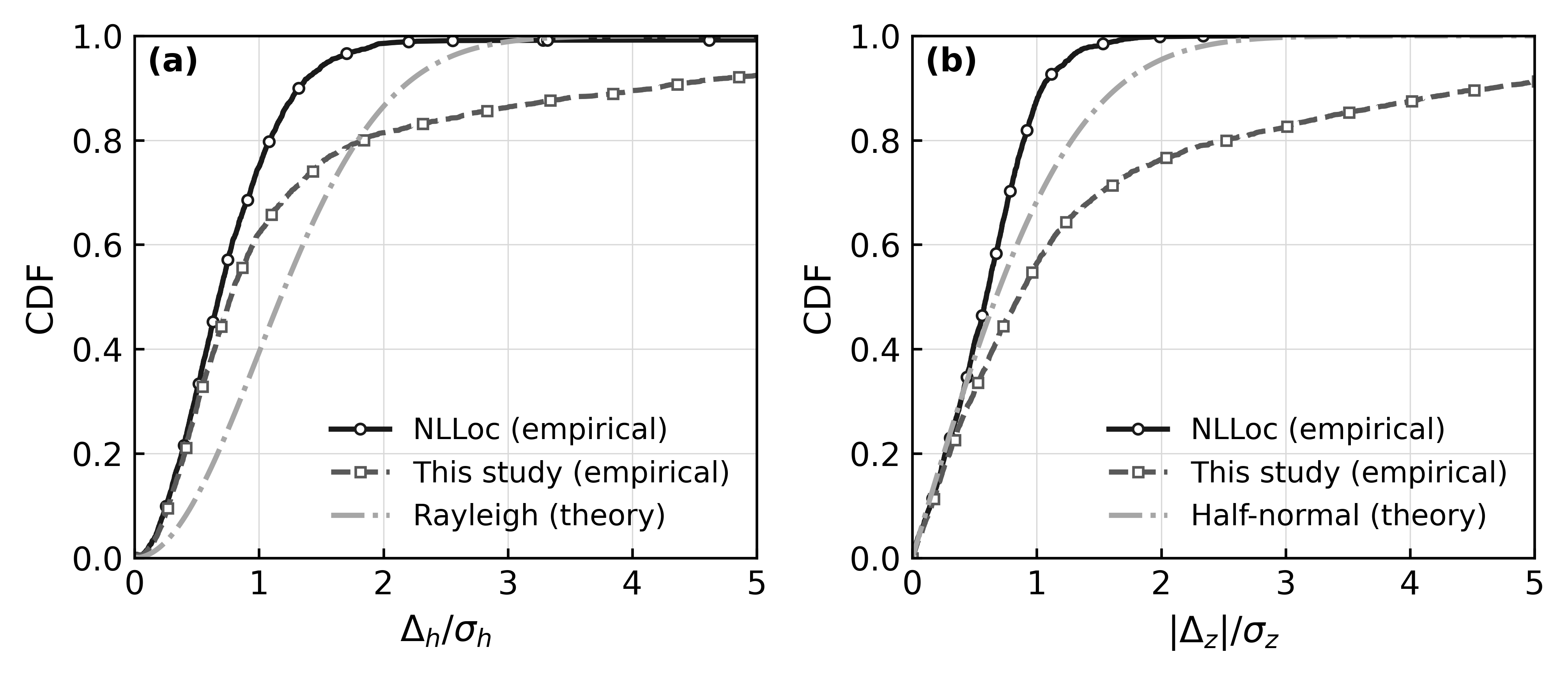}
\caption{Coverage tests for location uncertainty estimates. (a) Horizontal coverage shown as the empirical cumulative distribution of normalized horizontal errors $\Delta_h / \sigma_h$, compared with the theoretical Rayleigh distribution expected for a calibrated two-dimensional Gaussian posterior. (b) Depth coverage shown as the empirical cumulative distribution of normalized depth errors $|\Delta_z| / \sigma_z$, compared with the theoretical half-normal distribution expected for a calibrated one-dimensional Gaussian posterior.}
\label{fig:coverage}

\end{figure}

Figure~\ref{fig:coverage} presents empirical coverage curves for horizontal (2-D) and depth (1-D) location errors, normalized by the corresponding posterior uncertainties estimated from posterior samples. The coverage curves quantify the fraction of events whose true location lies within a given multiple of the reported uncertainty, and provide a diagnostic assessment of uncertainty behavior rather than a strict probabilistic calibration test.

It is important to note that a Student-$t$ likelihood is adopted in this study to model travel-time errors with heavy tails. As a result, the posterior distribution of earthquake location is generally non-Gaussian, even when the posterior uncertainty is summarized using second-order moments (standard deviations). Consequently, deviations from Gaussian-based reference distributions (Rayleigh for 2-D horizontal errors and half-normal for 1-D depth errors) should be interpreted as reflecting both uncertainty scaling and departures from Gaussianity in the posterior distribution.

For horizontal location errors (Figure~X, left), the empirical coverage curve of NLLoc rises more steeply than the Rayleigh reference at small normalized distances, indicating that the reported horizontal uncertainties tend to be under-dispersed, with true locations concentrated closer to the estimated epicenters than implied by the reported uncertainty scale. This behavior is consistent with the well-known tendency of linearized or locally Gaussian methods to underestimate uncertainty when strong nonlinearity or trade-offs are present.

In contrast, the proposed method exhibits a more gradual increase of coverage with increasing normalized horizontal error. The empirical curve lies closer to the Rayleigh reference over a broad range of $\Delta_h/\sigma_h$, indicating a more conservative and more stable representation of horizontal uncertainty. While the curve remains below the theoretical reference at large normalized distances, this behavior is expected under a heavy-tailed likelihood and reflects the influence of non-Gaussian posterior structure rather than systematic misestimation of uncertainty magnitude.

A similar pattern is observed for depth uncertainty (Figure~X, right). The NLLoc depth coverage curve increases rapidly and exceeds the half-normal reference at small normalized depth errors, again suggesting underestimation of depth uncertainty for a substantial fraction of events. In contrast, the proposed method produces a smoother and more gradual coverage curve that more closely follows the calibrated half-normal reference distribution.

Notably, the proposed method maintains consistently higher coverage at moderate normalized depth errors ($|\Delta z|/\sigma_z \sim 1$--$2$), indicating improved robustness of depth uncertainty estimates. This behavior is particularly important in regional-scale earthquake location, where depth trade-offs with origin time and velocity structure are common and often poorly resolved.

Taken together, these results suggest that the proposed method provides uncertainty estimates that are more conservative and more robust than those obtained from NLLoc, especially in regimes characterized by sparse station coverage, large azimuthal gaps, or long source--station distances. The improved uncertainty behavior can be attributed to the combined effects of a Student-$t$ likelihood, which mitigates the influence of outliers, and posterior sampling, which more fully captures nonlinearity and parameter trade-offs in the inverse problem.

It should be emphasized that the coverage curves are used here primarily for relative comparison between methods rather than as strict calibration diagnostics. Under a non-Gaussian posterior induced by a Student-$t$ likelihood, exact agreement with Gaussian reference distributions is neither expected nor required. Instead, the key result is the consistent improvement in uncertainty behavior relative to NLLoc, demonstrating that the proposed framework yields more reliable and interpretable uncertainty estimates for earthquake location.

\section{Posterior Travel-Time Uncertainty Characteristics}

\begin{figure}[htbp!]
\centering
\includegraphics[width=0.9\textwidth]{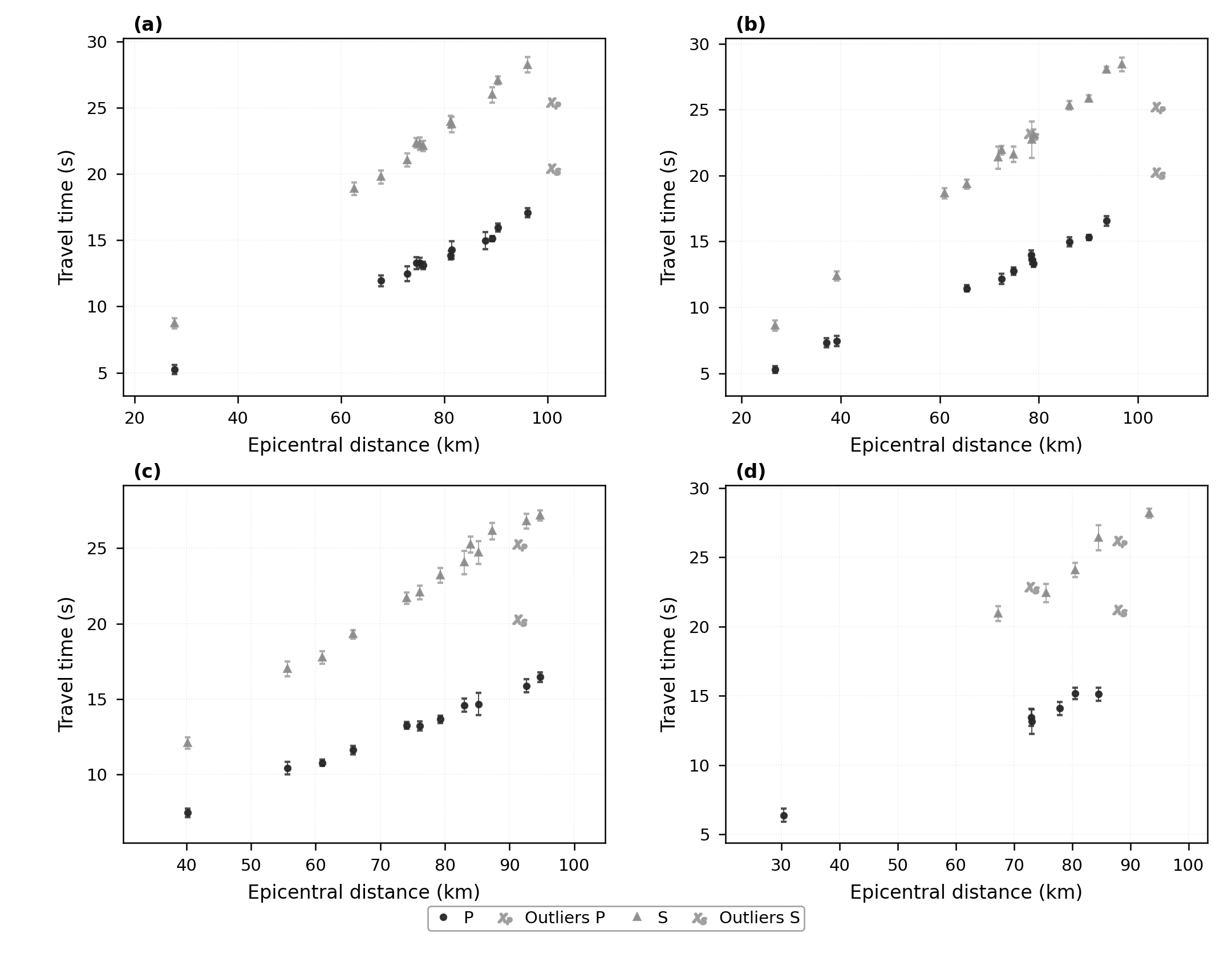}
\caption{
Posterior travel-time distributions as a function of epicentral distance for four representative earthquake events.
Panels (a)--(d) show P-phase (circles) and S-phase (triangles) travel-time samples obtained by propagating posterior hypocentral samples through the forward model.
Error bars indicate posterior uncertainty derived from the Student-$t$ likelihood formulation.
Across events, P-phase travel times exhibit systematically smaller uncertainties than S phases, while the overall spread varies substantially between events.
The differences among panels reflect event-dependent data coverage and source--path geometry, illustrating how posterior uncertainty is expressed directly in the data space.
}
\label{fig:dt_4events}
\end{figure}

To further illustrate how posterior uncertainty manifests in the data space, we examine the distribution of predicted travel times as a function of epicentral distance for four representative earthquake events.
Figure~\ref{fig:dt_4events} shows the posterior travel-time samples for both P and S phases, obtained by propagating the posterior hypocentral samples through the forward travel-time model.

Across all four events, the posterior travel-time samples form well-defined bands in the distance--time domain, with uncertainties that increase systematically with epicentral distance.
This behavior reflects the combined effects of velocity heterogeneity, ray-path geometry, and source-parameter trade-offs, and is consistent with expectations from regional-scale earthquake location problems.
The P-phase travel times exhibit relatively narrow posterior spreads, whereas the S-phase times show systematically larger uncertainty, reflecting their stronger sensitivity to shallow velocity structure and longer effective ray paths.

Notably, the posterior uncertainty is not uniform across events.
Panels (a) and (b) display compact and nearly linear travel-time trends, indicating that the arrival-time data provide strong constraints on both origin time and source location for these events.
In contrast, panels (c) and (d) show visibly broader posterior spreads, particularly for the S phases, suggesting weaker geometric constraints or stronger trade-offs between depth and origin time.
These differences highlight that posterior uncertainty is strongly event dependent and cannot be adequately characterized by a single global error metric.

From a Bayesian perspective, the spread of posterior travel-time samples directly represents the uncertainty implied by the data and the assumed likelihood model.
The Student-$t$ likelihood allows for heavier tails in the residual distribution, preventing individual outlying picks from dominating the inversion while still preserving the overall distance--time structure.
As a result, the posterior travel-time bands remain stable and physically interpretable even in cases where individual observations exhibit larger residuals.

We note that the degrees of freedom in the Student-$t$ likelihood are fixed rather than inferred.
Our objective here is robustness to outliers rather than exact likelihood calibration; 
a full hierarchical treatment of $\nu$ is left for future work.

Overall, these examples demonstrate that the proposed Bayesian framework yields coherent and data-consistent travel-time uncertainty estimates.
Rather than relying on linearized error propagation or uniform residual assumptions, the full posterior sampling provides a transparent description of how uncertainty varies with distance, phase type, and event geometry.

\bibliography{references.bib}
\end{document}